\documentclass{amsart}

\usepackage{graphicx}
\usepackage{amsfonts}
\usepackage{amsmath}
\usepackage{amssymb}

\newcommand{\func}[1]{\,\mathrm{#1}}
\newcommand{\aten}{\underline{\otimes}}

\newcommand{\nparagraph}[1]{\vskip.2cm\textit{#1}.}
\newcommand{\xparagraph}[1]{\textit{#1}.}

\newtheorem{theorem}{Theorem}[section]

\newtheorem{lemma}[theorem]{Lemma}

\theoremstyle{definition}
\newtheorem{definition}[theorem]{Definition}
\newtheorem{example}[theorem]{Example}

\theoremstyle{remark}
\newtheorem*{acknowledgements}{Acknowledgments}
\numberwithin{equation}{section}

\begin{document}

\title[Singular perturbations of QSDE]{Singular perturbation of quantum 
stochastic differential equations with coupling through an oscillator 
mode}

\author{John Gough}
\address{John Gough \\
	School of Computing and Informatics \\
	Nottingham Trent University,
	Nottingham, NG1 4BU, UK}
\email{john.gough@ntu.ac.uk}

\author{Ramon van Handel}
\address{Ramon van Handel \\
	Physical Measurement and Control 266-33 \\
                California Institute of Technology,
		Pasadena, CA 91125, USA}
\email{ramon@its.caltech.edu}
\thanks{R.v.H.\ is supported by the Army Research Office under Grant
W911NF-06-1-0378.}

\begin{abstract}
We consider a physical system which is coupled indirectly to a Markovian 
resevoir through an oscillator mode.  This is the case, for example, in 
the usual model of an atomic sample in a leaky optical cavity which is 
ubiquitous in quantum optics.  In the strong coupling limit the oscillator 
can be eliminated entirely from the model, leaving an effective direct 
coupling between the system and the resevoir.  Here we provide a 
mathematically rigorous treatment of this limit as a weak limit of the 
time evolution and observables on a suitably chosen exponential domain in 
Fock space.  The resulting effective model may contain emission and 
absorption as well as scattering interactions.
\end{abstract}

\keywords{Singular perturbation; Quantum stochastic differential 
equations; Hudson-Parthasarathy quantum stochastic calculus; Adiabatic 
elimination}

\maketitle

%%%%%%%%%%%%%%%%%%%%%%%%%%%%%%%%%%%%%%%%%%%%%%%%%%%%%%%%%%%%%%%%%%%%%%%%%

\section{Introduction}

\begin{figure}[t]
\includegraphics[width=\textwidth]{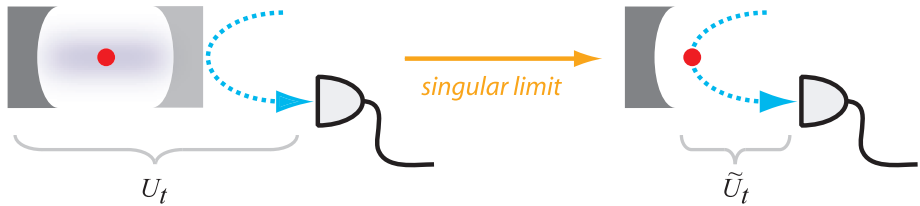}
\caption{\small Cartoon illustration of a prototypical problem that falls 
within the framework of this paper.  An atom is placed in a single-mode 
optical cavity.  One of the cavity mirrors is leaky, thus coupling the 
cavity mode to the external field (which may ultimately be detected).  In 
the ``bad cavity'' limit the mirror is made so transmissive that the 
cavity can be eliminated from the description of the model, leaving an 
effective direct interaction between the atom and the field.} 
\label{fig:singular} 
\end{figure}

The motivation for this article stems from the following problem in
quantum optics, illustrated in Fig.\ \ref{fig:singular}.  
Consider the canonical starting point of cavity QED, an atomic system in 
an optical cavity.  In many cases such a system is well modelled using 
only a single cavity mode.  The cavity is then effectively described by a 
single quantum harmonic oscillator, and the atom-cavity interaction 
Hamiltonian takes the form
\begin{equation}
\label{eq:hamiltinitial}
	H = 	E_{11}b^{\dag }b +
		E_{10}b^{\dag } +
		E_{01}b + E_{00},
\end{equation}
where $E_{ij}$ are operators acting on the atomic Hilbert space, 
$E_{ij}^\dag=E_{ji}$, and $b$, $b^\dag$ are the cavity mode annihilation 
and creation operators, respectively.  Usually one of the cavity mirrors 
is assumed to be perfectly reflective, while the other mirror allows some 
light to leak into the electromagnetic field outside the cavity and vice 
versa.  In the Markov approximation \cite{AFLu90,AGLu95}, the time 
evolution of the entire system (consisting of the atom, cavity and 
external field) is described by the unitary solution to the 
Hudson-Parthasarathy \cite{HP} quantum stochastic differential equation
\begin{equation}
\label{eq:qsdeinitial}
	dU_t=\left\{
		\sqrt{\gamma}\,b\,dA_t^\dag
		-\sqrt{\gamma}\,b^\dag\,dA_t
		-\frac{\gamma}{2}\,b^\dag b\,dt
		-iH\,dt
	\right\}U_t,
\end{equation}
where $A_t$, $A_t^\dag$ are the usual creation and annihilation processes 
in the external field.  The transmissivity of the leaky mirror is 
controlled by the positive constant $\gamma$.  

In many situations of practical interest, $\gamma$ will be quite large
compared to the strengths $\|E_{ij}\|$ of the atom-cavity interaction.  
When this is the case, one would expect that the presence of the cavity
has little qualitative influence on the atomic dynamics: the cavity is
then essentially transparent in the frequency range corresponding to the
atomic dynamics, so that the atoms ``see'' the external field directly.  
Similarly, we expect that measurements obtained from detection of the
outgoing field (e.g., by homodyne detection) would depend directly on the
atomic observables and would be essentially independent of the cavity
observables. The hope is, then, that the time evolution $U_t$ can be
described in some idealized limit by the unitary solution $\tilde U_t$ of
a new Hudson-Parthasarathy equation which involves only atomic operators
and the external field, and in which the cavity has been eliminated.  The
goal of this article is to make these ideas precise.

\nparagraph{Previous work}
The elimination of a leaky cavity in the bad cavity limit is an extremely
common procedure in the physics literature---so common, in fact, that most
papers state the resulting expression without further comment ({\it ``we
adiabatically eliminate the cavity, giving $\ldots$''}).  Often the
equation considered is a Lindblad-type master equation for the atom and
cavity; in our context, this (deterministic) differential equation for the
reduced density operator can be obtained by averaging over the field as in
\cite{HP}.  One method that is used to eliminate the cavity in such an
equation, see e.g.\ \cite{WiM93a}, involves expanding the density operator
in matrix elements, setting certain time derivatives to zero, then solving
algebraically to obtain an equation for the atomic matrix elements only.  
This method is commonly known as adiabatic elimination. Though such an
approach is not very rigorous, similar techniques can sometimes be
justified in the context of the classical theory of singular perturbations
(Tikhonov's theorem \cite{verhulst}).  A somewhat different approach, see
e.g.\ \cite{GaZ00}, uses projection operators and Laplace transform
techniques.  None of these techniques are applicable to the question posed
here, however, as we wish to retain the external field in the limiting
model. Hence we are seeking a singular perturbation result for quantum
stochastic differential equations, which is (to our knowledge) not yet
available in the literature.

A naive attempt at adiabatic elimination for quantum stochastic equations
is made in \cite{naive1} (see also \cite{naive2,naive3}).  These authors 
use the following procedure:
\begin{itemize}
\item First, they obtain Heisenberg equations of motion (in It\^o form) 
for the cavity annihilator $b_t=U_t^\dag bU_t$ and also for the relevant 
atomic operators.
\item Next, they set $\dot b_t=0$ (where the right-hand side is 
interpreted as ``quantum white noise'') and solve algebraically for $b_t$.
\item Next, they plug this expression into the atomic equations of motion.  
\item Finally, they interpret these equations as ``implicit'' equations
\cite{imexplicit} (a formal analog of Stratonovich equations) and 
convert to the ``explicit form'' (a formal analog of It\^o equations).  
The latter are considered to be the adiabatically eliminated Heisenberg 
equations of motion for the atomic operators.
\end{itemize}
Attempts at justifying this procedure run into a number of seemingly fatal 
problems.  Forgoing the issue of the mathematical well-posedness of 
``quantum white noise'', the approximation $\dot b_t=0$ seems incompatible 
with the fact that the right-hand side is formally infinite.  Next we have 
to deal with the interpretation of the resulting equations; even in the 
classical stochastic case, it is known that adiabatically eliminated 
expressions need not be of Stratonovich type (see \cite{gardiner} for some 
counterexamples); the singular limit is rather delicate and the resulting 
outcome depends on the way in which the limit is taken.  Ignoring even 
this issue, it should be pointed out that the implicit-explicit formalism 
introduced in \cite{imexplicit} (essentially along the lines of McShane's 
canonical extension \cite{mcshane,marcus}) does not even capture correctly 
the ordinary Markov limit in the presence of scattering interactions; 
compare the expressions in \cite{imexplicit} to the rigorous results 
obtained in \cite{Gou05}.  It is thus highly remarkable (if not 
miraculous!)\ that we can essentially reproduce the result of  
\cite{naive1} using the methods developed in this paper (see example 
\ref{ex:doherty} in section \ref{sec:limitdyn}).\footnote{
	We also mention \cite{warasz} where some results of \cite{naive1} 
	are reconsidered.  The results are only reproduced, however, at 
	the master equation level; in particular, the quantum noise 
	is not retained and the implicit-explicit formalism is not used in 
	those sections where results of \cite{naive1} are considered.
	Though the naive adiabatic elimination procedure used in
	\cite{naive1,naive2,naive3} is never well-justified, the 
	application of the implicit-explicit formalism is particularly 
	suspect in the presence of scattering interactions in view of the
	discrepancy between \cite{imexplicit} and \cite{Gou05}.  Note also
	that the manipulations in \cite{naive1} rely on the simple 
	commutation relations between position and momentum; they do not 
	work at all, e.g., if $E_{ab}$ are functions of angular momentum 
	operators.  It is thus quite surprising that, but unclear why, a 
	reasonable answer is obtained in the particular case considered in 
	\cite{naive1}.
}

\nparagraph{Statement of the problem}
Following \cite{gardiner}, we seek a {\it ``method by which fast variables
may be eliminated from the equations of motion in some well-defined
limit.''} Which limit to take is not entirely obvious at the outset; for
example, the naive choice $\gamma\to\infty$ only yields trivial results
(the cavity is forced to its ground state and the atomic dynamics
vanishes).  To define a nontrivial limit, we introduce the scaling
parameter $\varepsilon>0$ and make the substitution
$b\mapsto\varepsilon^{-1/2}b$ in (\ref{eq:hamiltinitial}) and
(\ref{eq:qsdeinitial}).  The limit $\varepsilon\to 0^+$ then has the
character of a central limit theorem, and provides a nontrivial result in
which the cavity is eliminated. (Note that similar scaling limits are used
in projection operator techniques for master equations \cite{GaZ00}.)

Our approach, then, is to proceed as follows.  First we make the above
substitution.  Next we switch to the interaction picture with respect to
the cavity-field interaction.  This gives rise to an interaction picture
time evolution in which the atom is driven by a quantum Ornstein-Uhlenbeck
process.  The limit $\varepsilon\to 0^+$ corresponds essentially to a
Markov limit of this equation, and consequently our proofs borrow heavily
from the methods developed to treat such limits (particularly from the
estimates developed in \cite{Gou05}).  However, our limits are of a
somewhat stronger character than those considered in
\cite{AFLu90,AGLu95,Gou05} as we take weak limits on a fixed domain in the
underlying Hilbert space, rather than ``limits in matrix elements'' where
the domain depends on $\varepsilon$. We also consider, aside from the time
evolution unitary and the Heisenberg evolution of the atomic observables,
the limiting behavior of the output field operators (which can be observed
e.g.\ through homodyne detection).

For concreteness, we will restrict ourselves to the model described by
(\ref{eq:hamiltinitial}) and (\ref{eq:qsdeinitial}).  This model is
already very rich and widely used in the literature in various scenarios.
Our results can also be extended to more complicated setups, in particular
to the case of multiple external fields and oscillators along the lines of
\cite{Gou06}; the subsequent extension to thermal and squeezed noises is
then also straightforward through the usual double Fock space
construction, see e.g.\ \cite{Gou03}, at least in the absence of 
scattering interactions ($E_{11}=0$).

\section{Preliminaries}

Throughout this article we work on the product Hilbert space
$\mathfrak{h}=\mathfrak{h}_{\mathrm{sys}}\otimes\mathfrak{h}_{\mathrm{osc}}
\otimes\mathfrak{h}_{\mathrm{resv}}$ consisting of a physical (e.g.\
atomic) system $\mathfrak{h}_{\mathrm{sys}}$, a quantum harmonic 
oscillator $\mathfrak{h}_{\mathrm{osc}}=\Gamma(\mathbb{C})$ (describing 
e.g.\ a cavity mode), and an external Bosonic resevoir
$\mathfrak{h}_{\mathrm{resv}}=\Gamma(L^2(\mathbb{R}_+))$ (describing e.g.\
the electromagnetic field).  Here $\Gamma(\mathfrak{h}')$ denotes the 
symmetric (Boson) Fock space over the one-particle Hilbert space 
$\mathfrak{h}'$.  We use the following notation for Fock space vectors: 
$|0\rangle\in\Gamma(\mathfrak{h}')$ denotes the vacuum vector, 
$|f\rangle\in\Gamma(\mathfrak{h}')$ denotes the exponential vector
corresponding to $f\in\mathfrak{h}'$, and 
$\mathcal{E}\subset\Gamma(\mathfrak{h}')$ denotes the linear space 
generated by the exponential vectors (the exponential domain).  We will 
also use the subscripts $|f\rangle_{\rm osc}$ or $|f\rangle_{\rm 
resv}$, and similarly $\mathcal{E}_{\rm osc}$, $\mathcal{E}_{\rm resv}$, 
wherever confusion may arise.

We define the following standard operators: $b$ and $b^\dag$ are the creation 
and annihilation operators on $\mathfrak{h}_{\mathrm{osc}}$, and $A_t$,
$A_t^\dag$ and $\Lambda_t$ are the usual annihilation, creation and gauge 
processes on $\mathfrak{h}_{\mathrm{resv}}$, respectively \cite{HP}.  We 
denote the ampliations of these operators to $\mathfrak{h}$ by the same 
symbols.  For any $f\in L^2(\mathbb{R}_+)$ and for any real, bounded $g\in 
L^\infty(\mathbb{R}_+)\cap L^2(\mathbb{R}_+)$ we also define the field 
operators \cite{HP}
$$
	A(f)=\int_0^\infty f(t)^*\,dA_t,\qquad
	A(f)^\dag = \int_0^\infty f(t)\,dA_t^\dag,\qquad
	\Lambda(g)=\int_0^\infty g(t)\,d\Lambda_t.
$$
We recall that the exponential domain $\mathcal{E}_{\rm resv}$ can be
extended to $\mathcal{E}_{\rm resv}'\supset\mathcal{E}_{\rm resv}$ in such
a way that $\mathcal{E}_{\rm resv}'$ is invariant under the action of
$A(f)$, $A(f)^\dag$ and $\Lambda(g)$, see e.g.\ \cite[pp.\ 61--65]{Mey93}.
Similarly $\mathcal{E}_{\rm osc}$ can be extended to $\mathcal{E}_{\rm
osc}'$ so that the latter is invariant under $b$, $b^\dag$ and $b^\dag b$.
This means in particular that the domain $\mathcal{D}'= \mathfrak{h}_{\rm
sys}\aten\mathcal{E}_{\rm osc}'\aten\mathcal{E}_{\rm
resv}'\subset\mathfrak{h}$ is invariant under finite linear combinations
of operators of the form $E\otimes\{b,b^\dag,b^\dag
b\}\otimes\{A(f),A(f)^\dag,\Lambda(g)\}$, and that commutators of such
operators are well defined on $\mathcal{D}'$.  Here $E$ is any bounded
operator on $\mathfrak{h}_{\rm sys}$ and $\aten$ denotes the algebraic
tensor product.  We recall also the useful identities $|\alpha\rangle_{\rm
osc}=\exp(\alpha b^{\dag})|0\rangle_{\rm osc}$ and
$|f\rangle_{\mathrm{resv}}=\exp(A(f)^{\dag}) |0\rangle_{\mathrm{resv}}$
\cite[sec.\ II.20]{Par92}.

The starting point for our investigation is the rescaled version of 
(\ref{eq:hamiltinitial}) and (\ref{eq:qsdeinitial}).  We consider the 
unitary solution $U_t(\varepsilon)$, given the initial condition 
$U_0(\varepsilon)=I$, to the Hudson-Parthasarathy quantum stochastic 
differential equation (QSDE) 
\begin{equation}
\label{eq:qsde}
	dU_{t}(\varepsilon) =\left\{ 
		\sqrt{\frac{\gamma}{\varepsilon}}\,
		b\,dA_{t}^{\dag}
		-\sqrt{\frac{\gamma}{\varepsilon}}\,
		b^{\dag}\,dA_{t}
		-\frac{\gamma}{2\varepsilon}\,b^{\dag}b\,dt
		-i\,H(\varepsilon)\,dt\right\}
		U_{t}(\varepsilon),
\end{equation}
where the Hamiltonian $H(\varepsilon) $ is taken to be of the form 
\begin{equation}
\label{eq:hamilt}
	H(\varepsilon) = \frac{1}{\varepsilon}\,E_{11}b^{\dag}b
	+
	\frac{1}{\sqrt{\varepsilon}}\,E_{10}b^{\dag}
	+
	\frac{1}{\sqrt{\varepsilon}}\,E_{01}b
	+E_{00}.
\end{equation}
Here $E_{ij}$, $\|E_{ij}\|<\infty$ are (the ampliations of) given bounded 
operators on $\mathfrak{h}_{\rm sys}$, $E_{ij}^{\dag}=E_{ji}$, and 
$\gamma,\varepsilon>0$ are positive constants.  The existence, uniqueness 
and unitarity of the solution of (\ref{eq:qsde}) are established in 
\cite{fagnola}.

\nparagraph{The interaction picture}
We are interested in the limit $\varepsilon\rightarrow 0^{+}$.  As the 
oscillator-resevoir dynamics becomes singular in this limit, the first 
step we take is to remove these dynamics by going over to the 
interaction representation.  To this end, define the oscillator-resevoir
time evolution $V_t(\varepsilon)$ as the unitary solution of
$$
	dV_{t}(\varepsilon) = 
	\left\{
		\sqrt{\frac{\gamma}{\varepsilon}}\,
		b\,dA_{t}^{\dag}
		-\sqrt{\frac{\gamma}{\varepsilon}}\,
		b^{\dag}\,dA_{t}
		-\frac{\gamma}{2\varepsilon}\,b^{\dag}b\,dt
	\right\}
	V_{t}(\varepsilon),
$$
where $V_0(\varepsilon)=I$.  Existence, uniqueness and unitarity are again 
guaranteed by \cite{fagnola}.  We wish to consider the unitary 
\begin{equation}
	\tilde{U}_{t}(\varepsilon) = V_{t}(\varepsilon)^\dag
	U_{t}(\varepsilon).
\end{equation}
Using the quantum It\^o rules \cite{HP}, we find that 
$\tilde{U}_{t}(\varepsilon)$ is given by the solution of the 
Schr{\"o}dinger equation
\begin{equation}
	\frac{d\tilde{U}_{t}(\varepsilon)}{dt} =
	-i\,\tilde{\Upsilon}_{t}(\varepsilon) 
	\tilde{U}_{t}(\varepsilon),\qquad \tilde{U}_0(\varepsilon)=I,
\end{equation}
with time-dependent interaction Hamiltonian 
\begin{equation}
	\tilde{\Upsilon}_{t}(\varepsilon) = V_{t}(\varepsilon)^{\dag}
	H(\varepsilon) V_{t}(\varepsilon).
\end{equation}
Note that $V_t(\varepsilon)$ commutes with any system operator $E$ (on 
$\mathfrak{h}_{\rm sys})$, so we have $E_t(\varepsilon)=U_t(\varepsilon)^\dag 
EU_t(\varepsilon)=\tilde U_t(\varepsilon)^\dag E\tilde U_t(\varepsilon)$.
Hence in order to study limits of the form $\lim_{\varepsilon\to 
0^+}E_t(\varepsilon)$ it is sufficient to consider $\tilde 
U_t(\varepsilon)$ rather than $U_t(\varepsilon)$.

\nparagraph{Quantum Ornstein-Uhlenbeck processes}
It is convenient to introduce the quantum Ornstein-Uhlenbeck (O-U)
annihilation and creation processes
\begin{equation}
	\tilde{a}_{t}(\varepsilon) =
	-\sqrt{\frac{\gamma}{4\varepsilon}}\,
	V_{t}(\varepsilon)^{\dag}bV_{t}(\varepsilon),
	\qquad
	\tilde{a}_{t}(\varepsilon)^\dag =
	-\sqrt{\frac{\gamma}{4\varepsilon}}\,
	V_{t}(\varepsilon)^{\dag}b^\dag V_{t}(\varepsilon).
\end{equation}
Using the quantum It\^o rules, we find that
$$
	d\tilde{a}_{t}(\varepsilon) =
	\frac{\gamma}{2\varepsilon}\,dA_t
	-\frac{\gamma}{2\varepsilon}\,\tilde{a}_{t}(\varepsilon)\,dt,
	\qquad
	\tilde{a}_{0}(\varepsilon) =
	-\sqrt{\frac{\gamma}{4\varepsilon}}\,b.
$$
Solving explicitly, we obtain
\begin{equation}
	\tilde{a}_{t}(\varepsilon) =
	\sqrt{\frac{\gamma}{4\varepsilon}}\,e^{-\gamma t/2\varepsilon }
	\left[ 
		\sqrt{\frac{\gamma}{\varepsilon}}
		\int_{0}^{t}e^{\gamma u/2\varepsilon}\,dA_{u} - b
	\right].
\end{equation}
This allows us to express the interaction Hamiltonian 
$\tilde{\Upsilon}_{t}(\varepsilon)$ in the form 
\begin{equation}
	\tilde{\Upsilon}_{t}(\varepsilon) =
	\frac{4}{\gamma}\,
	 E_{11}\tilde{a}_{t}(\varepsilon)^\dag
		\tilde{a}_{t}(\varepsilon) 
	-\frac{2}{\sqrt{\gamma}}\,
	E_{10}\tilde{a}_{t}(\varepsilon)^\dag 
	-\frac{2}{\sqrt{\gamma}}\,
	E_{01}\tilde{a}_{t}(\varepsilon) + E_{00}.
\end{equation}

\begin{lemma}
The O-U processes satisfy (on $\mathcal{D}'$) the commutation relations 
\begin{equation}
	[\tilde{a}_{t}(\varepsilon),\tilde{a}_{s}(\varepsilon)]=0,\qquad
	[\tilde{a}_{t}(\varepsilon),\tilde{a}_{s}(\varepsilon)^\dag] 
		= G_{\varepsilon}(t-s),
\end{equation}
where the correlation function $G_{\varepsilon}(\tau)$ is given by 
\begin{equation*}
	G_{\varepsilon}(\tau) =
		\frac{\gamma}{4\varepsilon}\,\exp\left(
		-\frac{\gamma|\tau|}{2\varepsilon}\right).
\end{equation*}
\end{lemma}

\begin{proof}
Recall \cite[sec.\ II.20]{Par92} the commutation relation on $\mathcal{D}'$
$$
	[A(f),A(g)^\dag]=\int_0^\infty f(t)^*g(t)\,dt.
$$
Hence we obtain
\begin{multline*}
	\left[
		e^{-\gamma t/2\varepsilon}
		\int_0^t e^{\gamma u/2\varepsilon}\,dA_u,
		e^{-\gamma s/2\varepsilon}
		\int_0^s e^{\gamma u/2\varepsilon}\,dA_u^\dag
	\right] = \\
	e^{-\gamma(t+s)/2\varepsilon}
	\int_0^{t\wedge s}e^{\gamma u/\varepsilon}\,du =
	\frac{\varepsilon}{\gamma}\,
	e^{-\gamma(t+s-2\,t\wedge s)/2\varepsilon}
	-\frac{\varepsilon}{\gamma}\,e^{-\gamma(t+s)/2\varepsilon},
\end{multline*}
where $t\wedge s=\min(t,s)$.  Now note that $t+s-2\,t\wedge s=|t-s|$.  
Hence writing out the full commutators and using $[b,b^\dag]=1$, the 
result follows. 
\end{proof}

The function $G_{\varepsilon}(\cdot) $ has the property of being strictly 
positive, symmetric and integrable with $\int_{-\infty}^{\infty}
G_{\varepsilon}(\tau)\,d\tau=1$. In the limit $\varepsilon\to 0^{+}$,
$G_{\varepsilon}(\cdot)$ therefore converges in the sense of distributions 
to a delta function at the origin. This means that the Ornstein-Uhlenbeck 
processes $\tilde{a}_{t}(\varepsilon)$, $\tilde{a}_{t}(\varepsilon)^\dag$
formally converge to quantum white noises as $\varepsilon \to 0^{+}$. Let 
us remark that these processes may now be written as 
\begin{equation}
\label{eq:convnoise}
	\tilde{a}_{t}(\varepsilon) = 
	2\int_{0}^{t} G_{\varepsilon}(t-s)\,dA_{s} -
	\sqrt{\frac{4\varepsilon}{\gamma}}\,G_{\varepsilon}(t)\,b.
\end{equation}

\section{Strong coupling limit ($\varepsilon\to 0^+$)}

For any $g\in L^2(\mathbb{R}_+)$ we define the future and past smoothed 
functions 
\begin{eqnarray*}
	g^{+}(t,\varepsilon) &=& 2\int_{0}^{\infty} g(t+\tau) 
		G_{\varepsilon}(\tau)\, d\tau
	= 2\int_t^\infty g(\tau) G_{\varepsilon}(t-\tau)\, d\tau, \\
	g^{-}(t,\varepsilon) &=& 2\int_{0}^{t} g(t-\tau)
		G_{\varepsilon}(\tau)\, d\tau
	= 2\int_0^t g(\tau) G_{\varepsilon}(t-\tau)\, d\tau.
\end{eqnarray*}
We will encounter such functions repeatedly in the following.  If $g$ were 
a continuous function, we would have the limits 
\begin{equation}
\label{eq:contilimit}
	\lim_{\varepsilon\to 0^{+}} g^{\pm}(t,\varepsilon)=g(t).
\end{equation}
The space $L^2(\mathbb{R}_+)$ is much too large, however, to ensure that 
the $\varepsilon\to 0^+$ limits of the smoothed functions are well 
behaved; consider for example a square integrable function with 
oscillatory discontinuity (e.g.\ $\sin(1/x)$).  To avoid such 
unpleasantness we will restrict our attention to the set of regulated 
square integrable functions, following \cite{GouCR}.

\begin{definition}
Let $L^2_\pm(\mathbb{R}_+)\subset L^2(\mathbb{R}_+)\cap 
L^\infty(\mathbb{R}_+)$ denote the set of square integrable bounded 
functions $f$ on the halfline such that the limits 
$f(t^\pm)=\lim_{s\to t^\pm}f(s)$ 
exist at every point $t\in\mathbb{R}_+$.  We denote by 
$\mathcal{E}^\pm_{\rm resv}\subset\mathcal{E}_{\rm resv}$ the restricted 
exponential domain generated by exponential vectors with amplitude 
functions in $L^2_\pm(\mathbb{R}_+)$.
\end{definition}

Before moving on, we make the following remarks:
\begin{enumerate}
\item Any $g\in L^2_\pm(\mathbb{R}_+)$ has at most a countable number of 
discontinuity points (see e.g.\ \cite[chapter 3]{Bil68}).  Hence for such 
$g$ Eq.\ (\ref{eq:contilimit}) holds for (Lebesgue-)a.e.\ $t\in\mathbb{R}_+$.
\item Note that if $g\in L^2_\pm(\mathbb{R}_+)$, then $\chi_{[0,s]}g\in 
L^2_\pm(\mathbb{R}_+)$ for any $s\in\mathbb{R}_+$.  Hence 
$\mathcal{E}^\pm_{\rm resv}$ is a suitable choice for the restricted 
exponential domain used in the construction of the Hudson-Parthasarathy 
stochastic integration theory \cite{HP}. 
\end{enumerate}

For future reference, we collect various $\varepsilon\to 0^+$ limits 
in the following lemma.

\begin{lemma}
For any $g\in L^2_\pm(\mathbb{R}_+)$, the following hold:
$g^\pm(t,\varepsilon)\xrightarrow{\varepsilon\to 0^+}g(t^\pm)$,
$$
	\lim_{\varepsilon\to 0^+}
	\int_0^\infty G_\varepsilon(t-s)g(s)\,ds
	=
	\lim_{\varepsilon\to 0^+}
	2\int_0^t G_\varepsilon(t-s)g^+(s,\varepsilon)\,ds
	=
	\frac{g(t^+)+g(t^-)}{2}.
$$
Moreover, all these expressions are equal to $g(t)$ for
(Lebesgue-)a.e.\ $t\in\mathbb{R}_+$; hence it follows that
$g^\pm(\cdot,\varepsilon)\xrightarrow[\varepsilon\to 0^+]{}g(\cdot)$
in $L^2(\mathbb{R}_+)$, etc.
\end{lemma}

\begin{proof}
The first statement follows from
$$
	g^+(t,\varepsilon)=\int_0^\infty
		\frac{\gamma}{2}\,e^{-\gamma\tau/2}\,
		g(t+\varepsilon\tau)\,d\tau
	\xrightarrow{\varepsilon\to 0^+}g(t^+),
$$
where we have used dominated convergence to take the limit.  Similarly
$$
	g^-(t,\varepsilon)=\int_0^\infty
		\frac{\gamma}{2}\,e^{-\gamma\tau/2}\,
		g(t-\varepsilon\tau)
		\chi_{[0,t/\varepsilon]}(\tau)
		\,d\tau
	\xrightarrow{\varepsilon\to 0^+}g(t^-).
$$
The third statement follows directly as
$$
	\int_0^\infty G_\varepsilon(t-s)g(s)\,ds=
	\frac{g^+(t,\varepsilon)+g^-(t,\varepsilon)}{2}.
$$
To prove the next statement, note that
$$
	2\int_0^t G_\varepsilon(t-s)g^+(s,\varepsilon)\,ds=
	\int_0^\infty g(\tau)\left[
		4\int_0^{t\wedge\tau}G_\varepsilon(t-s)
		G_\varepsilon(s-\tau)\,ds
	\right]d\tau.
$$
But straightforward calculation yields
$$
	4\int_0^{t\wedge\tau}G_\varepsilon(t-s)G_\varepsilon(s-\tau)\,ds
	=
	G_\varepsilon(t-\tau)-G_\varepsilon(\tau)\,\exp\left(
		-\frac{\gamma t}{2\varepsilon}
	\right),
$$
and the result follows directly.  Finally, the last statement of the lemma 
follows from the fact that $g$ has at most a countable number of 
discontinuities, together with the dominated convergence theorem. 
\end{proof}

For $f\in L_\pm^{2}(\mathbb{R}^{+})$, we define the smeared field operator 
\begin{equation}
	\tilde{A}(f,\varepsilon) =
	\int_{0}^{\infty} f(t)^{\ast}
	\tilde{a}_{t}(\varepsilon)\,dt.
\end{equation}
It is straightforward to obtain the commutation relation
\begin{equation*}
	[\tilde{A}(f,\varepsilon),\tilde{A}(g,\varepsilon)^{\dag}] 
	=\int_0^\infty f(t)^*G_{\varepsilon}g(t)\,dt
	=\frac{1}{2}\int_0^\infty f(t)^*(g^+(t,\varepsilon)+
		g^-(t,\varepsilon))\,dt,
\end{equation*}
where we have written $G_{\varepsilon}g(t) = \int_{0}^{\infty}
G_{\varepsilon}(t-s)g(s)\,ds=\frac{1}{2}g^{+}(t,\varepsilon)
+\frac{1}{2}g^{-}(t,\varepsilon)$.  Using (\ref{eq:convnoise}), we may 
also express the smeared field as
\begin{equation}
	\tilde{A}(f,\varepsilon) =
	A(f_{\varepsilon}^{+}) -
	\sqrt{\frac{\varepsilon}{\gamma}}\,
	f^{+}(0,\varepsilon)^{\ast}b,
\end{equation}
where $f_{\varepsilon}^{+}(t)=f^{+}(t,\varepsilon)$.  The second term 
ought to be negligible in the $\varepsilon\rightarrow 0^{+}$ limit and 
indeed, if $\varphi\in\mathfrak{h}_{\rm osc}$ is a vector with
$\|b\varphi\|<\infty$ and if $e(g)=|g\rangle_{\rm resv}$ is an exponential 
vector ($g\in L^2(\mathbb{R}_+)$), then 
\begin{multline*}
	\frac{\|(\tilde{A}(f,\varepsilon)-A(f))\,\varphi\otimes e(g)\|}
	{\|\varphi\otimes e(g)\|}
	\leq 
	\frac{\|A(f_\varepsilon^+ - f)\,e(g)\|}{\|e(g)\|}
	+\sqrt{\frac{\varepsilon}{\gamma}}\,f^{+}(0,\varepsilon)^{\ast}
	\frac{\|b\varphi\|}{\|\varphi\|} \\
	= \left|
		\int_0^\infty (f^+(t,\varepsilon)-f(t))^*g(t)\,dt
	\right|
	+\sqrt{\frac{\varepsilon}{\gamma}}\,f^{+}(0,\varepsilon)^{\ast}
        \frac{\|b\varphi\|}{\|\varphi\|}
	 \xrightarrow{\varepsilon\to 0^+} 0.
\end{multline*}
More generally, we can define the smeared Weyl operators 
\begin{equation}
	\tilde{W}(f,\varepsilon) =\exp\left\{\tilde{A}(f,\varepsilon)^{\dag}
		-\tilde{A}(f,\varepsilon)\right\},
\end{equation}
which satisfy the smeared canonical commutation relations
\begin{equation}
	\tilde{W}(f,\varepsilon) \tilde{W}(g,\varepsilon) =
	\tilde{W}(f+g,\varepsilon)\,\exp\left\{-i\,\func{Im}
	\int_0^\infty f(t)^*G_{\varepsilon}g(t)\,dt
	\right\}.
\end{equation}
The smeared Weyl operator $\tilde{W}(f,\varepsilon)$ ought to converge to 
the standard Weyl unitary $W(f)=I_{\rm osc}\otimes\exp\{A(f)^\dag-A(f)\}$
as $\varepsilon\to 0^+$.  It is indeed not difficult to establish that for 
arbitrary $f_{1},\cdots f_{n}\in L_\pm^{2}(\mathbb{R}_{+})$
\begin{equation*}
	\|(\tilde{W}(f_{1},\varepsilon) \cdots \tilde{W}(f_{n},\varepsilon)
	-W(f_{1})\cdots W(f_{n}))\,\varphi \otimes e(g)
	\| \xrightarrow{\varepsilon\to 0^+} 0.
\end{equation*}
This time, no restriction needs to be placed on $\varphi$.  This is a type 
of quantum central limit theorem, however, it is less abstract than the 
``limit in matrix elements'' traditionally encountered in the quantum 
probability literature \cite{AFLu90,AGLu95,Gou05} since the limit is taken 
on the fixed domain $\mathfrak{h}_{\rm osc}\aten\mathcal{E}_{\rm resv}$.  
The limiting operator is thus defined on the same Hilbert space, though it 
acts non-trivially on the noise space only.

\section{Limit dynamics}
\label{sec:limitdyn}

The limit of the process $\{\tilde{U}_{t}(\varepsilon):t\geq 0\}$ as
$\varepsilon\to 0^{+}$ is reminiscent of the Markov limits that have been
widely studied in mathematical physics \cite{AFLu90,AGLu95,Gou05}.  
Comparison with previous results suggests that the limit be again
described by a quantum stochastic process $\{\tilde{U}_{t}:t\geq 0\}$. We
wish to deduce this limiting process by studying the limit of matrix
elements $\langle\psi_{1}|\tilde{U}_{t}(\varepsilon)\psi_{2}\rangle$ for 
arbitrary vectors of the form
\begin{equation*}
	\psi_{i}=v_{i} \otimes |\alpha_{i}\rangle_{\rm osc}
	\otimes |f_{i}\rangle_{\rm resv},\qquad
	v_i\in\mathfrak{h}_{\rm sys},~\alpha_i\in\mathbb{C},~
	f_i\in L^2_\pm(\mathbb{R}_+).
\end{equation*}
In other words, we would like to obtain $\tilde U_t$ as the weak limit of 
$\tilde U_t(\varepsilon)$, as $\varepsilon\to 0^+$, on the domain
$\mathfrak{h}_{\rm sys}\aten\mathcal{E}_{\rm osc}\aten\mathcal{E}_{\rm 
resv}^\pm$.  We will similarly study weak limits of observables $\tilde 
U_t(\varepsilon)^\dag E\tilde U_t(\varepsilon)$, where $E$ is a system 
observable, on the same domain.  

Formally, we may expand $\tilde{U}_{t}(\varepsilon)$ as a Dyson series 
(by Picard iteration): 
\begin{equation}
	\tilde{U}_{t}(\varepsilon) = I +
	\sum_{n=1}^{\infty} (-i)^{n}
	\int_{\Delta_{n}(t)} ds_{n} \cdots ds_{1} \,
	\tilde\Upsilon_{s_{n}}(\varepsilon) \cdots 
	\tilde\Upsilon_{s_{1}}(\varepsilon),
\end{equation}
where the multi-time integrals are taken over the simplex 
\begin{equation*}
	\Delta_{n}(t) = \{ (s_{n},\dots,s_{1}) :
	t>s_{n}>\cdots>s_{1}>0\}.
\end{equation*}
The Dyson series expansion of $\langle\psi_{1}|\tilde{U}_{t}(\varepsilon)
\psi_{2}\rangle$ is given by
\begin{equation}
\label{eq:matrixdyson}
	\langle\psi_{1}|\psi_{2}\rangle
	+\sum_{n=1}^\infty (-i)^{n} \int_{\Delta_{n}(t)}
	ds_{n}\cdots ds_{1}\,\langle\psi_{1}
	|\tilde{\Upsilon}_{s_{n}}(\varepsilon)\cdots 
	\tilde{\Upsilon}_{s_{1}}(\varepsilon)\psi_{2}\rangle.
\end{equation}
The usual existence proof for differential equations by Picard iteration 
suggests that the Dyson series are convergent.  Following 
\cite{AFLu90,AGLu95,Gou05}, our basic approach will be as follows.  First, 
we obtain an estimate of the form
$$
	\left|\int_{\Delta_{n}(t)}
	ds_{n}\cdots ds_{1}\,\langle\psi_{1}
	|\tilde{\Upsilon}_{s_{n}}(\varepsilon)\cdots 
	\tilde{\Upsilon}_{s_{1}}(\varepsilon)\psi_{2}\rangle
	\right|\le \Omega(n),
$$
where $\Omega(n)$ is independent of $\varepsilon$ and such that
$\sum_{n=1}^\infty\Omega(n)<\infty$.  This establishes uniform convergence 
of the Dyson series (\ref{eq:matrixdyson}) on $\varepsilon>0$ (by the 
Weierstrass M-test).  Consequently, we may exchange the limit and the 
summation in the Dyson series: i.e., we have established that 
$\langle\psi_{1}|\tilde{U}_{t}(\varepsilon)\psi_{2}\rangle$ converges as 
$\varepsilon\to 0^+$ to
\begin{equation*}
	\langle\psi_{1}|\psi_{2}\rangle+\sum_{n=1}^\infty (-i)^{n} 
	\lim_{\varepsilon\to 0^+} \int_{\Delta_{n}(t)}
	ds_{n}\cdots ds_{1}\,\langle\psi_{1}
	|\tilde{\Upsilon}_{s_{n}}(\varepsilon)\cdots 
	\tilde{\Upsilon}_{s_{1}}(\varepsilon)\psi_{2}\rangle.
\end{equation*}
It then remains to determine the limiting form of every term in 
the Dyson series individually.  Summing these we obtain an (absolutely 
convergent) series expansion for the limiting matrix element, which we 
identify as the Dyson series expansion of the solution $\tilde U_t$ of a 
particular quantum stochastic differential equation.  This completes the 
proof.  Details can be found in section \ref{sec:proofs}.

In principle we should establish the results sketched above for every pair
of vectors $\psi_i$.  It is convenient, however, to reduce the problem to 
the study of the vacuum matrix element only.  To this end we use the 
following identity:
$$
	\psi_{i}  =
	\exp\{\alpha_{i}b^{\dag}\} 
	\exp\{A(f_{i})^\dag\}
	\, v_{i}\otimes |0\rangle_{\rm osc}\otimes|0\rangle_{\rm resv}.
$$
By commuting the operators $\exp\{\alpha_{i}b^{\dag}\}$ and
$\exp\{A(f_{i})^\dag\}$ past $\tilde{\Upsilon}_{s_{n}}(\varepsilon)
\cdots\tilde{\Upsilon}_{s_{1}}(\varepsilon)$ we can express the matrix
element in the integrand in terms of the vacuum, provided we make some 
simple modifications to $\tilde\Upsilon_t(\varepsilon)$. Hence we have to 
go through the proofs only once using the vectors 
$v_{i}\otimes|0\rangle_{\rm osc}\otimes|0\rangle_{\rm resv}$.

\begin{lemma}
\label{lem:commutethrough}
Define $\Phi=|0\rangle_{\rm osc}\otimes|0\rangle_{\rm resv}$.  Then
$$
	\langle\psi_{1}|\tilde{U}_{t}(\varepsilon)\psi_{2}\rangle =
	\langle v_{1}\otimes\Phi|\check{U}_{t}(\varepsilon)\,
	v_2\otimes\Phi\rangle\,
	\langle\alpha_1|\alpha_2\rangle_{\rm osc}\,
	\langle f_1|f_2\rangle_{\rm resv},
$$
where $\check{U}_{t}(\varepsilon)$ is the modification of 
$\tilde{U}_{t}(\varepsilon)$ obtained by the replacements
\begin{eqnarray*}
	\tilde{a}_{t}(\varepsilon) & \longmapsto &
	\check{a}_{t}^{-}(\varepsilon) =
	\tilde{a}_{t}(\varepsilon)
	+f_{2}^-(t,\varepsilon)
	-\sqrt{\frac{4\varepsilon}{\gamma}}\,G_{\varepsilon}(t)\,\alpha_{2},
	\\
	\tilde{a}_{t}(\varepsilon)^{\dag} & \longmapsto &
	\check{a}_{t}^{+}(\varepsilon) =
	\tilde{a}_{t}(\varepsilon)^{\dag}
	+f_{1}^-(t,\varepsilon)^{\ast}
	-\sqrt{\frac{4\varepsilon}{\gamma}}\,G_{\varepsilon}(t)\,
	\alpha_{1}^*.
\end{eqnarray*}
\end{lemma}

\begin{proof}
Key here are the simple identities (on $\mathcal{D}'$)
$$
	be^{\alpha b^{\dag}}=e^{\alpha b^{\dag}}(b+\alpha),\qquad
	A(g)e^{A(f)^\dag}=e^{A(f)^\dag}\left(
	A(g)+\int_0^\infty g(s)^*f(s)\,ds\right),
$$
which allow us to write using (\ref{eq:convnoise})
$$
	\tilde a_t(\varepsilon)
	e^{\alpha b^{\dag}} 
	e^{A(f)^\dag}=
	e^{\alpha b^{\dag}} 
	e^{A(f)^\dag}\left(
		\tilde a_t(\varepsilon)+f^-(t,\varepsilon)
		-\sqrt{\frac{4\varepsilon}{\gamma}}\,G_\varepsilon(t)\,
	\alpha\right).
$$
Hence starting from any term in the Dyson series of the form
$$
	\langle\psi_{1}
	|\tilde{\Upsilon}_{s_{n}}(\varepsilon)\cdots 
	\tilde{\Upsilon}_{s_{1}}(\varepsilon)\psi_{2}\rangle
	=
	\langle v_1\otimes\Phi|e^{\alpha_1^*b}e^{A(f_1)}
	\tilde{\Upsilon}_{s_{n}}(\varepsilon)\cdots 
	\tilde{\Upsilon}_{s_{1}}(\varepsilon)
	e^{A(f_2)^\dag} e^{\alpha_2 b^{\dag}}
	\,v_2\otimes\Phi\rangle,
$$
the result follows using the above relations if we use additionally that
$$
	e^{\alpha_1^*b}e^{A(f_1)}e^{A(f_2)^\dag} e^{\alpha_2 b^{\dag}}
	=e^{A(f_2)^\dag} e^{\alpha_2 b^{\dag}}e^{\alpha_1^*b}e^{A(f_1)}\,
	\langle\alpha_1|\alpha_2\rangle_{\rm osc}\,
	\langle f_1|f_2\rangle_{\rm resv}
$$
and that $e^{\alpha_i^*b}e^{A(f_i)}\,\Phi=\Phi$.
\end{proof}

The Dyson expansion for the matrix element 
$\langle\psi_{1}|\tilde{U}_{t}(\varepsilon)\psi_{2}\rangle$ may now be 
written, up to a constant prefactor of 
$\langle\alpha_1|\alpha_2\rangle_{\rm osc}\,\langle f_1|f_2\rangle_{\rm 
resv}$, as 
\begin{equation*}
	\langle v_{1}|v_{2}\rangle
	+\sum_{n=1}^\infty(-i)^{n}\int_{\Delta_{n}(t)}
	ds_{n}\cdots ds_{1}\,\langle v_{1}\otimes\Phi |
	\check{\Upsilon}_{s_{n}}(\varepsilon) \cdots 
	\check{\Upsilon}_{s_{1}}(\varepsilon)\,
	v_{2}\otimes\Phi\rangle.
\end{equation*}
Here $\check{\Upsilon}_{t}(\varepsilon)$ is obtained from 
$\tilde{\Upsilon}_{t}(\varepsilon)$ by making the translations above, 
that is, 
\begin{equation}
	\check{\Upsilon}_{t}(\varepsilon) =
	\frac{4}{\gamma}\,
	 E_{11}\check{a}_{t}^+(\varepsilon)
		\check{a}_{t}^-(\varepsilon) 
	-\frac{2}{\sqrt{\gamma}}\,
	E_{10}\check{a}_{t}^+(\varepsilon) 
	-\frac{2}{\sqrt{\gamma}}\,
	E_{01}\check{a}_{t}^-(\varepsilon) + E_{00}.
\end{equation}
As the new processes $\check{a}_{t}^{\pm}(\varepsilon)$ are linear in the 
original O-U processes, we may write 
\begin{equation*}
	\check{\Upsilon}_{t}(\varepsilon)
	=\check{E}_{11}(t,\varepsilon) 
	\tilde{a}_{t}(\varepsilon)^{\dag}\tilde{a}_{t}(\varepsilon) 
	+\check{E}_{10}(t,\varepsilon)\tilde{a}_{t}(\varepsilon)^{\dag}
	+\check{E}_{01}(t,\varepsilon)\tilde{a}_{t}(\varepsilon) 
	+\check{E}_{00}(t,\varepsilon).
\end{equation*}
The coefficients $\check{E}_{ij}(t,\varepsilon)$ are easily worked out, 
however, our main interest will be in their limit values: we have
for (Lebesgue-)a.e.\ $t\in\mathbb{R}_+$
\begin{equation*}
\begin{split}
	\check{E}_{11}(t,\varepsilon) =& \,
	\check{E}_{11}(t) = 
	\frac{4}{\gamma}\,E_{11}, \\
	\lim_{\varepsilon\rightarrow 0^{+}}\check{E}_{10}(t,\varepsilon)
		=&\, \check{E}_{10}(t) = -\frac{2}{\sqrt{\gamma}}\,E_{10} 
			+ \frac{4}{\gamma}\,E_{11}f_{2}(t), \\
	\lim_{\varepsilon\rightarrow 0^{+}}\check{E}_{01}(t,\varepsilon)
		=&\, \check{E}_{01}(t) = -\frac{2}{\sqrt{\gamma}}\,E_{01}
			+ \frac{4}{\gamma}\,f_{1}(t)^{\ast }E_{11}, \\
	\lim_{\varepsilon\rightarrow 0^{+}}\check{E}_{00}(t,\varepsilon)
		=&\, \check{E}_{00}(t) = 
			E_{00} - \frac{2}{\sqrt{\gamma}}\,E_{01}f_{2}(t) 
			- \frac{2}{\sqrt{\gamma}}\,f_{1}(t)^{\ast}E_{10}
			+ \frac{4}{\gamma}\,f_{1}(t)^{\ast}E_{11}f_{2}(t),
\end{split}
\end{equation*}
the limits being uniform in the strong topology. Note that these limits 
depend only on the functions $f_{i}$ describing the resevoir: the 
parameters $\alpha_{i}$ for the oscillator have disappeared, indicating 
that the oscillator is indeed eliminated as $\varepsilon\to 0^+$.

The expansion of $\langle\psi_{1}|\tilde{U}_{t}(\varepsilon)\psi_{2}
\rangle/\langle\alpha_1|\alpha_2\rangle_{\rm osc}\,\langle f_1|f_2
\rangle_{\rm resv}$ may now be written as 
\begin{eqnarray*}
	\langle v_1|v_2\rangle~+
	&&\sum_{n=1}^\infty(-i)^{n}\int_{\Delta_{n}(t)}ds_{n}\cdots ds_{1} \\
	&&\times \sum_{\alpha_{n}\beta_{n}}\cdots 
		\sum_{\alpha_{1}\beta_{1}}
	\langle v_{1}|\check{E}_{\alpha_{n}\beta_{n}}(s_{n},\varepsilon) 
	\cdots \check{E}_{\alpha_{1}\beta_{1}}(s_{1},\varepsilon)
	\,v_{2}\rangle  \\
	&&\times \langle\Phi|
	[\tilde{a}_{s_{n}}(\varepsilon)^{\dag}]^{\alpha_{n}}
	[\tilde{a}_{s_{n}}(\varepsilon)]^{\beta_{n}}
	\cdots
	[\tilde{a}_{s_{1}}(\varepsilon)^{\dag}]^{\alpha_{1}}
	[\tilde{a}_{s_{1}}(\varepsilon)]^{\beta_{1}}\,\Phi\rangle,
\end{eqnarray*}
where $\alpha_k,\beta_k$ are summed over the values $0,1$ and we write
$[x]^{0}=1$, $[x]^{1}=x$.  It is this form of the expansion that will be 
most useful in the proofs (section \ref{sec:proofs}).

We can now state the main results.

\begin{theorem}
\label{thm:mainunitary}
Suppose the system operators $E_{ij}$ are bounded with 
$\|E_{11}\|<\gamma/2$. Then there exists a unitary quantum stochastic 
process $\{\tilde{U}_{t}:t\geq 0\}$ such that
\begin{equation*}
	\lim_{\varepsilon\rightarrow 0^{+}}
	\langle\psi_{1}|\tilde{U}_{t}(\varepsilon)\psi_{2}\rangle =
	\langle\psi_{1}|\tilde{U}_{t}\psi_{2}\rangle 
\end{equation*}
for any pair of vectors $\psi_{1,2}\in\mathfrak{h}_{\rm 
sys}\aten\mathcal{E}_{\rm osc}\aten\mathcal{E}_{\rm resv}^\pm$.
The process $\{\tilde{U}_{t}:t\geq 0\}$ satisfies a QSDE of 
Hudson-Parthasarathy type
\begin{equation*}
	d\tilde{U}_{t}=\left\{ 
		(\tilde W-I)\,d\Lambda_{t}+\tilde L\,dA_{t}^{\dag}
		-\tilde L^\dag \tilde W\,dA_{t}
		-\frac{1}{2}\tilde L^\dag \tilde L\,dt
		-i\tilde H\,dt
	\right\} \tilde{U}_{t},\qquad
	\tilde{U}_{0}=I,
\end{equation*}
where the coefficients are given by the expressions
$$
	\tilde W=\frac{\gamma/2-iE_{11}}{\gamma/2+iE_{11}},~~~~
	\tilde L=\frac{i\sqrt{\gamma}}{\gamma/2+iE_{11}}\,E_{10},~~~~
	\tilde H=E_{00}+{\rm Im}\left\{
		E_{01}(\gamma/2+iE_{11})^{-1}E_{10}
	\right\}.
$$
\end{theorem}

\begin{theorem}
\label{thm:mainatom}
Under the conditions of Thm.\ \ref{thm:mainunitary}, we have 
convergence of the Heisenberg evolution: for any bounded operator $E$ 
on $\mathfrak{h}_{\rm sys}$ and $\psi_{i}\in\mathfrak{h}_{\rm
sys}\aten\mathcal{E}_{\rm osc}\aten\mathcal{E}_{\rm resv}^\pm$
\begin{equation*}
	\lim_{\varepsilon\rightarrow 0^{+}}
	\langle\psi_{1}|{U}_{t}(\varepsilon)^\dag E
	{U}_{t}(\varepsilon)\psi_{2}\rangle =
	\lim_{\varepsilon\rightarrow 0^{+}}
	\langle\psi_{1}|\tilde{U}_{t}(\varepsilon)^\dag E
	\tilde{U}_{t}(\varepsilon)\psi_{2}\rangle =
	\langle\psi_{1}|\tilde{U}_{t}^\dag E\tilde{U}_t\psi_{2}\rangle.
\end{equation*}
\end{theorem}

Let us demonstrate these results for some models used in the physics 
literature.

\begin{example}
\label{ex:doherty}
Doherty {\it et al.}\ \cite{naive1} consider the following system, in our 
notation:
$$
	\gamma=2\kappa,\qquad
	H=E_{00}-\frac{g_0^2}{\Delta}\,\cos^2(k_Lx)\,b^\dag b,
$$
where $x$ is the atomic position operator on $\mathfrak{h}_{\rm 
sys}=L^2(\mathbb{R})$ and $E_{00}$ is a free Hamiltonian.\footnote{
	Technically their Hamiltonian $E_{00}=p_x^2/2m$ is unbounded,
	but we sweep this under the rug.
}
According to Theorem \ref{thm:mainunitary}, the limiting time evolution is 
given by
$$
	d\tilde U_t=\left\{
		(\tilde W-I)\,d\Lambda_t-iE_{00}\,dt
	\right\}\tilde U_t,
	\qquad
	\tilde W=\frac{\kappa+ig_0^2\cos^2(k_Lx)/\Delta}
	{\kappa-ig_0^2\cos^2(k_Lx)/\Delta},
$$
provided that $\|E_{11}\|=g_0^2/\Delta<\kappa$.  According to Theorem 
\ref{thm:mainatom} and the quantum It\^o rules, the limiting Heisenberg 
evolution of an atomic operator $X$ is given by
$$
	d(\tilde U_t^\dag X\tilde U_t)=
	\tilde U_t^\dag\left(i[E_{00},X]\,dt+
	(\tilde W^\dag X\tilde W-X)\,d\Lambda_t\right)\tilde U_t.
$$
Compare this expression to Eq.\ (2.16ab) in \cite{naive1}, taking into 
account the identity $\exp\{2i\,\tan^{-1}(x)\}=
(\cos\{\tan^{-1}(x)\}+i\,\sin\{\tan^{-1}(x)\})^2=
(1+ix)/(1-ix)$.
\end{example}

\begin{example}
The following interaction Hamiltonian is often used to describe the 
coupling between a collection of atomic spins (total spin $J$, i.e.\ 
$\mathfrak{h}_{\rm sys}=\mathbb{C}^{2J+1}$) and a far detuned driven 
cavity mode (see e.g.\ \cite{Tho02}):
$$
	H = \chi F_z b^\dag b+\mathcal{U}(b^\dag+b) + E_{00}.
$$
Here $E_{00}$ is a free atomic Hamiltonian and $\chi,\mathcal{U}$ are 
real constants.  By Theorem \ref{thm:mainunitary}, the operators $\tilde 
W$, $\tilde L$, $\tilde H$ in the limiting QSDE become
$$
	\tilde W=\frac{\gamma/2-i\chi F_z}{\gamma/2+i\chi F_z},\quad
	\tilde L=\frac{i\mathcal{U}\sqrt{\gamma}}{\gamma/2+i\chi F_z},
	\quad
	\tilde H=E_{00}-\frac{\chi\mathcal{U}^2F_z}{\gamma^2/4+
		\chi^2F_z^2},
$$
provided $\|E_{11}\|=\chi J<\gamma/2$.  A common assumption in the 
literature (a reasonable one if the adiabatic approximation is good) is 
that $\|2\chi F_z/\gamma\|=2\chi J/\gamma\ll 1$; the conventional 
adiabatically eliminated master equation (such as the one used in 
\cite{Tho02}) is now recovered by calculating the master equation
corresponding to $\tilde U_t$, then expanding $\tilde L$ and
$\tilde H$ to first order with respect to $2\chi F_z/\gamma$.
\end{example}

\section{Limit output fields}
\label{sec:output}

Aside from the limit dynamics of the system observables, as in
Theorem \ref{thm:mainatom}, we are also interested in the limiting
behavior of the resevoir observables after interaction with the system and
oscillator. In optical systems, for example, these observables can be
detected (using, e.g., homodyne detection \cite{Bar03}) and the observed
photocurrent can be used for statistical inference of the unmeasured
system observables (quantum filtering theory \cite{Bel92b,BHJ06}). The
behavior of these observables in the singular limit is thus of significant
interest for the modelling of quantum measurements.

To investigate the limit of the field observables we study the 
convergence of matrix elements of the form $\langle\psi_1|U_t(\varepsilon)^\dag 
W(g_{t]})U_t(\varepsilon)\,\psi_2\rangle$, where $\psi_{1,2}\in
\mathfrak{h}_{\rm sys}\aten\mathcal{E}_{\rm osc}\aten\mathcal{E}_{\rm 
resv}^\pm$, $W(g)$ is the usual Weyl operator with $g\in 
L^2_\pm(\mathbb{R}_+)$ and $g_{t]}(s)=g(s)\chi_{[0,t]}(s)$.  Note that 
unlike in the system operator case, $V_t(\varepsilon)$ does not commute 
with $W(g_{t]})$.  However, we obtain using the quantum It\^o rules
$$
	V_t(\varepsilon)^\dag W(g_{t]})V_t(\varepsilon) =
	\exp\left\{
		A(g_{t]})^\dag-A(g_{t]})
		-2(\tilde A(g_{t]},\varepsilon)^\dag
		-\tilde A(g_{t]},\varepsilon))
	\right\},
$$
or, expressing this in terms of the usual Weyl operators,
$$
	V_t(\varepsilon)^\dag W(g_{t]})V_t(\varepsilon) =
	W(g_{t]}-2g_{t],\varepsilon}^+)\,
	\exp\left\{
		\sqrt{\frac{4\varepsilon}{\gamma}}\,
		\left(g_{t]}^+(0,\varepsilon)b^\dag-
		g_{t]}^+(0,\varepsilon)^*b\right)
	\right\}.
$$
Hence we can write
\begin{equation}
\label{eq:resevoirexpect}
	U_t(\varepsilon)^\dag W(g_{t]}) U_t(\varepsilon) =
	\tilde U_t(\varepsilon)^\dag 
		W(g_{t]}-2g_{t],\varepsilon}^+)
		B(\sqrt{4\varepsilon/\gamma}\,g_{t]}^+(0,\varepsilon))
	\tilde U_t(\varepsilon),
\end{equation}
where denote by $B(\alpha)=\exp\{\alpha b^\dag-\alpha^*b\}$ the 
Weyl operator for the oscillator.  Using the Dyson series for
$\tilde U_t(\varepsilon)$, we now expand
$\langle\psi_1|U_t(\varepsilon)^\dag W(g_{t]}) U_t(\varepsilon)\psi_2\rangle$ 
as
\begin{eqnarray*}
	&& \sum_{n=0}^\infty\sum_{m=0}^\infty
 	(-i)^{n-m} 
	\int_{\Delta_{m}(t)} dt_{m}\cdots dt_{1}
	\int_{\Delta_{n}(t)} ds_{n}\cdots ds_{1} \\
	&& \times\langle\psi_{1}|
	\tilde{\Upsilon}_{t_{1}}(\varepsilon)\cdots 
	\tilde{\Upsilon}_{t_{m}}(\varepsilon)
		W(g_{t]}-2g_{t],\varepsilon}^+)
		B(\sqrt{4\varepsilon/\gamma}\,g_{t]}^+(0,\varepsilon))
	\tilde{\Upsilon}_{s_{n}}(\varepsilon)\cdots 
	\tilde{\Upsilon}_{s_{1}}(\varepsilon)
	\psi_{2}\rangle
\end{eqnarray*}
(for notational simplicity we have used the convention 
$\int_{\Delta_0(t)}\cdots=I$ here).
The limit of this expression is most easily studied by commuting the Weyl 
operators through the $\tilde{\Upsilon}_{s}(\varepsilon)$ terms, in the 
spirit of lemma \ref{lem:commutethrough}.  In particular, using
$$
	W(f)=\exp\{A(f)^\dag\}\exp\{-A(f)\}\exp\left\{
		-\frac{1}{2}\int_0^\infty |f(t)|^2\,dt
	\right\},
$$
and similarly
$$
	B(\alpha)=\exp\{\alpha b^\dag\}\exp\{-\alpha^*b\}\,
	\exp\{-|\alpha|^2/2\},
$$
then moving the conjugated terms to the left and the remaining terms to 
the right (where they operate trivially on the vacuum), the problem can be 
reduced to the manipulations used in the proof of Theorem \ref{thm:mainatom}.
Details can be found in section \ref{sec:proofs}.

We can already guess at this point, however, what the answer should be. As
$g_{t]}(s)-2g_{t]}^+(s,\varepsilon)\to -g_{t]}(s)$ $s$-a.e., and as
$\sqrt{4\varepsilon/\gamma}\,g_{t]}^+(0,\varepsilon)\to 0$, we expect that
$$
	\tilde U_t(\varepsilon)^\dag W(g_{t]}-2g_{t],\varepsilon}^+)
	B(\sqrt{4\varepsilon/\gamma}\,g_{t]}^+(0,\varepsilon))\tilde 
	U_t(\varepsilon) \xrightarrow{\varepsilon\to 0^+}
	\tilde U_t^\dag W(-g_{t]})\tilde U_t.
$$
This is in fact the case.

\begin{theorem}
\label{thm:mainfield}
Under the conditions of Theorem \ref{thm:mainunitary}, we have the 
following: for any $\psi_{1,2}\in\mathfrak{h}_{\rm
sys}\aten\mathcal{E}_{\rm osc}\aten\mathcal{E}_{\rm resv}^\pm$ and
$g\in L^2_\pm(\mathbb{R}_+)$
\begin{equation*}
	\lim_{\varepsilon\rightarrow 0^{+}}
	\langle\psi_{1}|{U}_{t}(\varepsilon)^\dag W(g_{t]})
	{U}_{t}(\varepsilon)\psi_{2}\rangle =
	\langle\psi_{1}|\tilde{U}_{t}^\dag 
	W(-g_{t]})\tilde{U}_t\psi_{2}\rangle.
\end{equation*}
\end{theorem}

\section{Proofs}
\label{sec:proofs}

In the previous sections we have set up the problems to be solved, and we 
have investigated in detail the Ornstein-Uhlenbeck noises and the 
associated correlation functions and limits.  With this preliminary spade 
work at hand, the remaining (technical) part of the proofs, as outlined in 
section \ref{sec:limitdyn}, follows to a large extent from the proofs and 
estimates in \cite{Gou05}.  Below we work through the required steps in 
the proofs, however we refer to \cite{Gou05} for some detailed 
calculations.
\newline

\subsection{Proof of Theorem \ref{thm:mainunitary}}~\newline
\xparagraph{Wick ordering} All steps of the proofs require us to evaluate 
the matrix elements
$$
	\langle\Phi|
	[\tilde{a}_{s_{n}}(\varepsilon)^{\dag}]^{\alpha_{n}}
	[\tilde{a}_{s_{n}}(\varepsilon)]^{\beta_{n}}
	\cdots
	[\tilde{a}_{s_{1}}(\varepsilon)^{\dag}]^{\alpha_{1}}
	[\tilde{a}_{s_{1}}(\varepsilon)]^{\beta_{1}}\,\Phi\rangle
$$
that appear in the Dyson series.  The solution to this problem is well 
known and proceeds by applying Wick's lemma \cite{Schweber}.  For given 
sequences $\alpha=(\alpha_i)$, $\beta=(\beta_i)$, define the sets
$P(\alpha)=\{i:\alpha_i=1\}$ and $Q(\beta)=\{i:\beta_i=1\}$.  Let 
$\mathfrak{J}(\alpha,\beta)$ be the set of all maps $J:P(\alpha)\to 
Q(\beta)$ that are bijections and that are increasing, i.e.\ $J(i)>i$.  
Then by Wick's lemma we obtain
$$
	\langle\Phi|
	[\tilde{a}_{s_{n}}(\varepsilon)^{\dag}]^{\alpha_{n}}
	[\tilde{a}_{s_{n}}(\varepsilon)]^{\beta_{n}}
	\cdots
	[\tilde{a}_{s_{1}}(\varepsilon)^{\dag}]^{\alpha_{1}}
	[\tilde{a}_{s_{1}}(\varepsilon)]^{\beta_{1}}\,\Phi\rangle
	= \sum_{J\in\mathfrak{J}(\alpha,\beta)}\prod_{i\in P(\alpha)}
		G_\varepsilon(s_{J(i)}-s_i).
$$
Diagrammatically, this can be represented as follows.  Write $n$ vertices 
on a line:
\vskip.25cm
\begin{center}
\setlength{\unitlength}{.05cm}
\begin{picture}(120,20)
\put(20,10){\circle*{2}}
\put(30,10){\circle*{2}}
\put(40,10){\circle*{2}}
\put(50,10){\circle*{2}}
\put(60,10){\circle*{2}}
\put(70,10){\circle*{2}}
\put(80,10){\circle*{2}}
\put(90,10){\circle*{2}}
\put(100,10){\circle*{2}}
\put(18,1){$s_{n}$}
\put(88,1){$s_{2}$}
\put(98,1){$s_{1}$}
\put(58,1){$s_{j}$}
\put(10,10){\dashbox{0.5}(100,0){ }}
\end{picture}
\end{center}
\vskip.25cm
\noindent
For each vertex $j$, draw ingoing and outcoming lines corresponding to the 
values of $\alpha_j$ and $\beta_j$, as follows:
\vskip.25cm
\begin{center}
\begin{tabular}{cccc}
{\small $\alpha_j=\beta_j=1$} & 
{\small $\alpha_j=1$, $\beta_j=0$} &
{\small $\alpha_j=0$, $\beta_j=1$} &
{\small $\alpha_j=\beta_j=0$}
\\
{\tiny ~} & {\tiny ~} & {\tiny ~} & {\tiny ~}
\\
\setlength{\unitlength}{.1cm}
\begin{picture}(10,5)
\put(0,0){\dashbox{0.5}(10,0){ }}
\thicklines
\put(5,0){\circle*{1}}
\put(0,0){\oval(10,10)[tr]}
\put(10,0){\oval(10,10)[tl]}
\end{picture}
&
\setlength{\unitlength}{.1cm}
\begin{picture}(10,5)
\put(0,0){\dashbox{0.5}(10,0){ }}
\thicklines
\put(5,0){\circle*{1}}
\put(0,0){\oval(10,10)[tr]}
\end{picture}
&
\setlength{\unitlength}{.1cm}
\begin{picture}(10,5)
\put(0,0){\dashbox{0.5}(10,0){ }}
\thicklines
\put(5,0){\circle*{1}}
\put(10,0){\oval(10,10)[tl]}
\end{picture}
&
\setlength{\unitlength}{.1cm}
\begin{picture}(10,5)
\put(0,0){\dashbox{0.5}(10,0){ }}
\thicklines
\put(5,0){\circle*{1}}
\end{picture}
\\
{\small $~~s_j$} &
{\small $~~s_j$} &
{\small $~~s_j$} &
{\small $~~s_j$}
\end{tabular}
\end{center}
\vskip.25cm
\noindent
Note that $P(\alpha)$ is the set of vertices that have outgoing lines, 
whereas $Q(\beta)$ is the set of vertices that have incoming lines.
Next, connect every outgoing line to one of the incoming lines at a later 
time (i.e.\ form pair contractions), in such a way that all the lines are 
connected to exactly one other line.  For example:
\vskip.25cm
\begin{center}
\setlength{\unitlength}{.05cm}
\begin{picture}(120,20)
\put(20,10){\circle*{2}}
\put(30,10){\circle*{2}}
\put(40,10){\circle*{2}}
\put(50,10){\circle*{2}}
\put(60,10){\circle*{2}}
\put(70,10){\circle*{2}}
\put(80,10){\circle*{2}}
\put(90,10){\circle*{2}}
\put(100,10){\circle*{2}}
\put(30,10){\oval(20,20)[t]}
\put(40,10){\oval(20,20)[t]}
\put(60,10){\oval(20,20)[t]}
\put(90,10){\oval(20,20)[t]}
\put(18,1){$s_{9}$}
\put(28,1){$s_{8}$}
\put(38,1){$s_{7}$}
\put(48,1){$s_{6}$}
\put(58,1){$s_{5}$}
\put(68,1){$s_{4}$}
\put(78,1){$s_{3}$}
\put(88,1){$s_{2}$}
\put(98,1){$s_{1}$}
\put(10,10){\dashbox{0.5}(100,0){ }}
\end{picture}
\end{center}
%\vskip.25cm
\noindent
The ways in which this can be done are in one-to-one correspondence with 
the elements of $\mathfrak{J}(\alpha,\beta)$; the contracted vertices are 
then simply the pairs $(i,J(i))$ where $i\in P(\alpha)$.  Wick's lemma
tells us that the sum over all such (Goldstone) diagrams gives precisely 
the vacuum matrix element we are seeking.

\nparagraph{Step 1: a uniform estimate}  Our first goal is to find a 
uniform (in $\varepsilon$) estimate on every term $M_\varepsilon(n)$ in 
the Dyson series:
\begin{multline*}
M_\varepsilon(n) = \left|
	\int_{\Delta_{n}(t)}ds_{n}\cdots ds_{1}
	\sum_{\alpha_{n}\beta_{n}}\cdots
                \sum_{\alpha_{1}\beta_{1}}
	\langle v_{1}|\check{E}_{\alpha_{n}\beta_{n}}(s_{n},\varepsilon) 
	\cdots \check{E}_{\alpha_{1}\beta_{1}}(s_{1},\varepsilon)
	\,v_{2}\rangle  
\right.
\\
\left.\phantom{\sum_{\beta_{1}}\int}
	\times \langle\Phi|
	[\tilde{a}_{s_{n}}(\varepsilon)^{\dag}]^{\alpha_{n}}
	[\tilde{a}_{s_{n}}(\varepsilon)]^{\beta_{n}}
	\cdots
	[\tilde{a}_{s_{1}}(\varepsilon)^{\dag}]^{\alpha_{1}}
	[\tilde{a}_{s_{1}}(\varepsilon)]^{\beta_{1}}\,\Phi\rangle
\right|.
\end{multline*}
Note that in this expression there is a summation over $\alpha$ and 
$\beta$.  Hence every $n$-vertex Goldstone diagram is going to appear in 
the sum when we apply Wick's lemma, not just those with fixed 
incoming/outcoming lines for each vertex (as for fixed $\alpha$, $\beta$).
Whenever this is the case it is convenient, rather than first summing over 
$\mathfrak{J}(\alpha,\beta)$ and then over $\alpha,\beta$, to arrange the 
sum in a slightly different way.

Every $n$-vertex Goldstone diagram can be described completely by
specifying a partition of the set $\{1,\ldots,n\}$; each part of the
partition corresponds to a group of vertices that are connected.  For
example, the nine-vertex example diagram above corresponds to the
partition $\{\{1,3\},\{2\},\{4,6,8\},\{5\},\{7,9\}\}$.  The corresponding
values of $\alpha$ and $\beta$ are easily reconstructed: a singleton
vertex has $\alpha=\beta=0$, and for a doubleton or higher the first
vertex has $\alpha=1$, $\beta=0$, the last vertex has $\alpha=0$,
$\beta=1$, and the vertices in the middle have $\alpha=\beta=1$.  The sum
over $\alpha,\beta$ and $\mathfrak{J}(\alpha,\beta)$, which appears in the
expression for $M_\varepsilon(n)$ after applying Wick's lemma, can now be
replaced by the sum over all partitions $\mathfrak{B}_n$ of the $n$-point
set.

We need to refine the summation a little further.  To every partition we
associate a sequence ${\bf n}=(n_j)_{j\in\mathbb{N}}$ of integers, where
$n_j$ counts the number of $j$-tuples that make up the partition (e.g.,
the example diagram above has $\mathbf{n}= (2,2,1,0,0,0,\ldots)$).  We
denote by $E({\bf n})=\sum_jjn_j$ the number of vertices in the partition
(so $E({\bf n})=n$ for a partition of $n$ vertices) and by $N({\bf
n})=\sum_jn_j$ the number of parts that make up the partition.  Of course
there are many partitions that have the same occupation sequence ${\bf
n}$; the set of all such partitions is denoted $\mathfrak{B}_{\bf
n}\subset\mathfrak{B}_{E({\bf n})}$.  Summing over $\mathfrak{B}_n$ now
corresponds to summing first over $\mathfrak{B}_{\bf n}$, then over all
${\bf n}$ with $E({\bf n})=n$.

We are now ready to bound $M_\varepsilon(n)$.  First, note that
$$
	|\langle v_{1}|\check{E}_{\alpha_{n}\beta_{n}}(s_{n},\varepsilon) 
	\cdots \check{E}_{\alpha_{1}\beta_{1}}(s_{1},\varepsilon)
	\,v_{2}\rangle|
	\le \|v_1\|\,\|v_2\|\,C_{\alpha_n\beta_n}\cdots C_{\alpha_1\beta_1},
$$
where $C_{\alpha\beta}$ are finite positive constants that depend only on 
$\|E_{\alpha\beta}\|$, $\max_{[0,t]}f_{1,2}$ and $\gamma$.  In particular, 
$C_{11}=(4/\gamma)\|E_{11}\|$ and we will write $C=\max_{\alpha\beta}
C_{\alpha\beta}$.  For any $\alpha,\beta$ corresponding to the occupation 
sequence ${\bf n}$, the number of times that $\alpha=1$, $\beta=1$ will be 
$\sum_{j>2}(j-2)n_{j}=E(\mathbf{n})-N(\mathbf{n})+n_{1}$.  Hence
$$
	C_{\alpha_{n}\beta_{n}}\cdots C_{\alpha_{1}\beta_{1}}\leq
	C_{11}^{E(\mathbf{n})-N(\mathbf{n})+n_{1}}
		C^{N(\mathbf{n})-n_{1}}.
$$
We can thus estimate $M_\varepsilon(n)/\|v_1\|\,\|v_2\|$ by
$$
	\sum_{\bf n}^{E({\bf n})=n}
	C_{11}^{E(\mathbf{n})-N(\mathbf{n})+n_{1}} C^{N(\mathbf{n})-n_{1}}
	\sum_{\rho\in\mathfrak{B}_{\bf n}}
	\int_{\Delta_n(t)}ds_n\cdots ds_1\,
	\prod_{(i,j)\sim\rho} G_\varepsilon(s_i-s_j),
$$
where $(i,j)\sim\rho$ denotes that the vertices $i$ and $j$ are 
contracted in the partition $\rho$.  A clever argument due to Pul{\'e} can 
now be extended to show that
$$
	\sum_{\rho\in\mathfrak{B}_{\bf n}}
	\int_{\Delta_n(t)}ds_n\cdots ds_1\,
	\prod_{(i,j)\sim\rho} G_\varepsilon(s_i-s_j) \le
	\frac{1}{n_{1}!n_{2}!\cdots}\,
	\frac{t^{N(\mathbf{n})}}{2^{E(\mathbf{n})-N(\mathbf{n})}}.
$$
Essentially, the trick is to rewrite the sum over $\mathfrak{B}_{\bf n}$ 
of integrals over the simplex as a single integral over a union of 
simplices, which can then be estimated; see \cite[section 7]{Gou05} 
for details.  If $C_{11}>0$, we obtain the following estimate 
uniformly in $\varepsilon$:
$$
	M_\varepsilon(n) \le
	\Omega(n)=
	\|v_1\|\,\|v_2\|
	\sum_{\bf n}^{E({\bf n})=n}
	\frac{
		e^{A E({\bf n})+B N({\bf n})}
	}{
		n_1!n_2!\cdots
	},
$$
where $A=\log(C_{11}/2)$ and $B=\log(t\vee 1)+\log(C^2\vee 1)+
\log(C_{11}^{-2}\vee 1)+\log 2$.  Summing over $n$, we obtain
$$
	\frac{1}{\|v_1\|\,\|v_2\|}\sum_n\Omega(n)=\sum_{\bf n}
	\frac{
		e^{A E({\bf n})+B N({\bf n})}
	}{
		n_1!n_2!\cdots
	}=
	\prod_{k=1}^\infty\sum_{n=0}^\infty
	\frac{e^{(kA+B)n}}{n!}=\exp\left\{
		\frac{e^{A+B}}{1-e^A}
	\right\},
$$
provided that $e^A=C_{11}/2<1$, i.e.\ the sum converges provided that
$\|E_{11}\|<\gamma/2$.  Recall that this was a condition of Theorem
\ref{thm:mainunitary}.  If $C_{11}=0$ we obtain a slightly different 
estimate, which is however even simpler to sum (most terms vanish).

Now that we have a uniform estimate, the Weierstrass M-test guarantees 
that the Dyson series converges uniformly in $\varepsilon$.  Consequently, 
we can calculate the limit of the Dyson series as $\varepsilon\to 0^+$ 
simply by calculating the limit of each diagram independently, then 
summing all these terms.  This is what we will do below.

\nparagraph{Step 2: principal terms in the Dyson series}  The contribution 
of a 
single Goldstone diagram to the Dyson series has the form
$$
	\int_{\Delta_{n}(t)}ds_{n}\cdots ds_{1}
	\langle v_{1}|\check{E}_{\alpha_{n}\beta_{n}}(s_{n},\varepsilon) 
	\cdots \check{E}_{\alpha_{1}\beta_{1}}(s_{1},\varepsilon)
	\,v_{2}\rangle  
	\,\prod_{i\in P(\alpha)}G_\varepsilon(s_{J(i)}-s_i)
$$
for some $J\in\mathfrak{J}(\alpha,\beta)$.  A diagram will be called 
time-consecutive if $J(i)=i+1$ for every $i\in P(\alpha)$.  We claim that 
in the limit $\varepsilon\to 0^+$ any diagram that is not time-consecutive 
vanishes: hence we only need to retain time-consecutive diagrams.

To see this, first note that the magnitude of the diagram above is 
bounded by
$$
	C^n\,\|v_1\|\,\|v_2\|
	\int_{\Delta_{n}(t)}ds_{n}\cdots ds_{1}
	\,\prod_{i\in P(\alpha)}G_\varepsilon(s_{J(i)}-s_i).
$$
The limit of the latter integral is not difficult to evaluate explicitly.  
In particular, if $J$ is not time-consecutive then the integral vanishes 
in the limit $\varepsilon\to 0^+$.  For example, suppose that $J(i)\ne 
i+1$, so that $s_{J(i)}>s_{i+1}$ a.e.\ in $\Delta_{n}(t)$.  Then
$$
	\int_0^{s_{i+1}}ds_i\,
	G_\varepsilon(s_{J(i)}-s_i)
	\xrightarrow{\varepsilon\to 0^+}
	0~~\mbox{for any }s_{J(i)}>s_{i+1}
$$
by dominated convergence, as $G_\varepsilon(s_{J(i)}-s_i)$ is uniformly 
bounded on $[0,s_{i+1}]$ whenever $s_{J(i)}>s_{i+1}$ and 
$G_\varepsilon(s_{J(i)}-s_i)\to 0$ pointwise.  On the other hand,
$$
	\int_0^{s_{i+1}}ds_i\,
	G_\varepsilon(s_{J(i)}-s_i) \le
	\int_{-\infty}^{s_{J(i)}}ds_i\,
	G_\varepsilon(s_{J(i)}-s_i)=
	\frac{1}{2}~~\mbox{for any }s_{J(i)}\ge s_{i+1}.
$$
Hence we have by dominated convergence
$$
	\int_0^{s_{J(i)+1}}ds_{J(i)}\cdots\int_0^{s_{i+1}}ds_i\,
	G_\varepsilon(s_{J(i)}-s_i)
	\xrightarrow{\varepsilon\to 0^+}
	0.
$$
Proceeding in the same way, we can show that any diagram that is not 
time-consecutive vanishes as $\varepsilon\to 0^+$ \cite[lemma 6.1]{Gou05}.

It remains to consider the time-consecutive diagrams, for example:
\vskip.25cm
\begin{center}
\setlength{\unitlength}{.05cm}
\begin{picture}(120,20)
\put(20,10){\circle*{2}}
\put(30,10){\circle*{2}}
\put(40,10){\circle*{2}}
\put(50,10){\circle*{2}}
\put(60,10){\circle*{2}}
\put(70,10){\circle*{2}}
\put(80,10){\circle*{2}}
\put(90,10){\circle*{2}}
\put(100,10){\circle*{2}}
\put(25,10){\oval(10,10)[t]}
\put(35,10){\oval(10,10)[t]}
\put(55,10){\oval(10,10)[t]}
\put(85,10){\oval(10,10)[t]}
\put(95,10){\oval(10,10)[t]}
\put(18,1){$s_{9}$}
\put(28,1){$s_{8}$}
\put(38,1){$s_{7}$}
\put(48,1){$s_{6}$}
\put(58,1){$s_{5}$}
\put(68,1){$s_{4}$}
\put(78,1){$s_{3}$}
\put(88,1){$s_{2}$}
\put(98,1){$s_{1}$}
\put(10,10){\dashbox{0.5}(100,0){ }}
\end{picture}
\end{center}
%\vskip.25cm
\noindent
These diagrams have a particularly simple structure: any
such diagram is uniquely described by listing, in increasing time order, 
the number of vertices in each connected component.  For example, the 
diagram above is described by the sequence $(3,1,2,3)$.  In this way, any
$n$-vertex diagram with $m$ connected components is described by a set of
integers $r_1,\ldots,r_m$ such that $r_1+\cdots+r_m=n$.  Now suppose that 
$J\in\mathfrak{J}(\alpha,\beta)$ is a time-consecutive diagram that is 
described by the sequence $r_1,\ldots,r_m$ with $r_1+\cdots+r_m=n$.  It 
is not difficult to verify that
\begin{multline*}
	\int_{\Delta_{n}(t)}ds_{n}\cdots ds_{1}
	\,\Xi(s_1,\ldots,s_n)
	\prod_{i\in P(\alpha)}G_\varepsilon(s_{J(i)}-s_i)
	\xrightarrow{\varepsilon\to 0^+}
	\\
	\frac{1}{2^{n-m}}
	\int_{\Delta_{m}(t)}dt_{m}\cdots dt_{1}
	\,\Xi(
		\underbrace{t_1,t_1,\ldots,t_1}_{r_1\mbox{ times}},
		\underbrace{t_2,t_2,\ldots,t_2}_{r_2\mbox{ times}},
		\ldots,
		\underbrace{t_m,t_m,\ldots,t_m}_{r_m\mbox{ times}}
	)
\end{multline*}
for any function $\Xi\in L^2_\pm(\mathbb{R}_+^n)$.  Note that $n-m$ is 
precisely the number of contractions in the diagram $r_1,\ldots,r_m$.

\nparagraph{Step 3: resumming the Dyson series} 
We now compose the various steps made thus far.  Starting from the $n$th 
term in the Dyson expansion, using Wick's lemma, retaining only the 
time-consecutive terms, and taking the limit as $\varepsilon\to 0^+$ gives
\begin{multline*}
	\int_{\Delta_{n}(t)}ds_{n}\cdots ds_{1}
	\sum_{\alpha_{n}\beta_{n}}\cdots
                \sum_{\alpha_{1}\beta_{1}}
	\langle v_{1}|\check{E}_{\alpha_{n}\beta_{n}}(s_{n},\varepsilon) 
	\cdots \check{E}_{\alpha_{1}\beta_{1}}(s_{1},\varepsilon)
	\,v_{2}\rangle  
\\
	\times \langle\Phi|
	[\tilde{a}_{s_{n}}(\varepsilon)^{\dag}]^{\alpha_{n}}
	[\tilde{a}_{s_{n}}(\varepsilon)]^{\beta_{n}}
	\cdots
	[\tilde{a}_{s_{1}}(\varepsilon)^{\dag}]^{\alpha_{1}}
	[\tilde{a}_{s_{1}}(\varepsilon)]^{\beta_{1}}\,\Phi\rangle
\\
	\xrightarrow{\varepsilon\to 0^+}
	\sum_{m}\sum_{r_1,\ldots,r_m\ge 1}^{r_1+\cdots+r_m=n}
	\frac{1}{2^{n-m}}
	\int_{\Delta_{m}(t)}dt_{m}\cdots dt_{1}\,
	\langle v_{1}|\check{E}^{(r_m)}(t_{m}) 
	\cdots \check{E}^{(r_1)}(t_{1})
	\,v_{2}\rangle,
\end{multline*}
where we have written
$$
	\check{E}^{(r)}(t)=\left\{
	\begin{array}{ll}
		\check{E}_{00}(t) & \quad r=1, \\
		\check{E}_{01}(t)(\check{E}_{11}(t))^{r-2}
		\check{E}_{10}(t) & \quad r\ge 2.
	\end{array}
	\right.
$$
Let us now sum all the terms in the limiting Dyson series: this gives
\begin{multline*}
	\langle v_1,v_2\rangle+
	\sum_{n=1}^\infty (-i)^n
	\sum_{m}\sum_{r_1,\ldots,r_m\ge 1}^{r_1+\cdots+r_m=n}
	\frac{1}{2^{n-m}} \\ \times
	\int_{\Delta_{m}(t)}dt_{m}\cdots dt_{1}\,
	\langle v_{1}|\check{E}^{(r_m)}(t_{m}) 
	\cdots \check{E}^{(r_1)}(t_{1})
	\,v_{2}\rangle.
\end{multline*}
Now use the fact that $n-m=\sum_k (r_k-1)$ to rewrite this expression as
$$
	\langle v_1,v_2\rangle+
	\sum_{m}
	\int_{\Delta_{m}(t)}dt_{m}\cdots dt_{1}\,
	\langle v_{1}|
	\left(\sum_{r_m\ge 1}\frac{\check{E}^{(r_m)}(t_{m})}{i^{r_m}2^{r_m-1}}
	\right)
	\cdots 
	\left(\sum_{r_1\ge 1}\frac{\check{E}^{(r_1)}(t_{1})}{i^{r_1}2^{r_1-1}}
	\right)
	\,v_{2}\rangle.
$$
But note that we can sum
\begin{multline*}
	\sum_{r\ge 1}\frac{\check{E}^{(r)}(t)}{i^{r}2^{r-1}}=
	-i\check{E}_{00}(t)
	-\frac{1}{2}\check{E}_{01}(t)
	\left(\sum_{r\ge 0}\frac{(\check{E}_{11}(t))^r}{(2i)^r}\right)
	\check{E}_{10}(t)
\\
	=
	-i\check{E}_{00}(t)
	-\frac{1}{2}\check{E}_{01}(t)
	\,\frac{1}{1+i\check{E}_{11}(t)/2}\,
	\check{E}_{10}(t)
\end{multline*}
provided that $\|i\check{E}_{11}(t)/2\|=(2/\gamma)\|E_{11}\|<1$, which was 
already required for uniform convergence of the Dyson series.
Finally we define
$$
	L_{\alpha\beta} =
	\left[-i\,E_{\alpha\beta}-E_{\alpha 1}\,
	\frac{1}{\gamma/2+iE_{11}}\,E_{1\beta}\right]
	\left(
		-\frac{2}{\sqrt{\gamma}}
	\right)^{\alpha+\beta},
$$
and note that we can write
$$
	-i\check{E}_{00}(t)
	-\frac{1}{2}\check{E}_{01}(t)
	\,\frac{1}{1+i\check{E}_{11}(t)/2}\,
	\check{E}_{10}(t)=
	\sum_{\alpha\beta}[f_1(t)^*]^\alpha L_{\alpha\beta}
		[f_2(t)]^\beta.
$$
Hence the Dyson expansion for $\langle\psi_{1}|\tilde{U}_{t}(\varepsilon)
\psi_{2}\rangle/\langle\alpha_1|\alpha_2\rangle_{\rm osc}\,\langle f_1|f_2
\rangle_{\rm resv}$ may be written, in the limit $\varepsilon\to 0^+$, as 
\begin{multline*}
	\langle v_1,v_2\rangle+\sum_m
	\sum_{\alpha_{m}\beta_{m}}\cdots \sum_{\alpha_{1}\beta_{1}}
	\langle v_{1}|L_{\alpha_{m}\beta_{m}}\cdots L_{\alpha_{1}\beta_{1}}
	\,v_{2}\rangle \\
	\times \int_{\Delta_{m}(t)} dt_{m}\cdots dt_{1}\,
	[f_1(t_m)^*]^{\alpha_m}[f_2(t_m)]^{\beta_m}
	\cdots
	[f_1(t_1)^*]^{\alpha_1}[f_2(t_1)]^{\beta_1}.
\end{multline*}

\nparagraph{Step 4: the limit unitary}
It remains to investigate the relation of the limiting Dyson series given 
above to the unitary evolution $\tilde U_t$.  Consider a 
Hudson-Parthasarathy equation of the form
$$
	d\tilde U_t=\left\{
		L_{11}\,d\Lambda_t+
		L_{10}\,dA_t^\dag+
		L_{01}\,dA_t+
		L_{00}\,dt
	\right\}\tilde U_t.
$$
By Picard iteration, the solution $\tilde U_t$ can be developed into its 
chaos expansion 
$$
	\tilde U_t=I+\sum_m
	\sum_{\alpha_{m}\beta_{m}}\cdots \sum_{\alpha_{1}\beta_{1}}
	\int_{\Delta_m(t)} L_{\alpha_{m}\beta_{m}}\cdots L_{\alpha_{1}\beta_{1}}
	\,d\Lambda_{t_m}^{\alpha_{m}\beta_{m}}\cdots
	d\Lambda_{t_1}^{\alpha_{1}\beta_{1}},
$$
where we have used the Evans notation $\Lambda^{11}_t=\Lambda_t$, 
$\Lambda^{10}_t=A_t^\dag$, $\Lambda^{01}_t=A_t$, $\Lambda^{00}_t=t$
(see e.g.\ \cite[page 151]{Mey93}).  Using the usual formula for the 
matrix elements of stochastic integrals, it is evident that 
$\langle\psi_1|\tilde U_t\,\psi_2\rangle$ coincides with the limiting 
Dyson series above.  It remains to notice, as is verified through 
straightforward manipulations, that $L_{11}=\tilde W-I$, $L_{10}=\tilde 
L$, $L_{01}=-\tilde L^\dag\tilde W$, and $L_{00}=-i\tilde H-\tilde L^\dag 
\tilde L/2$.  The proof of Theorem \ref{thm:mainunitary} is complete.

\subsection{Proof of Theorem \ref{thm:mainatom}}

Conceptually, little changes when we are interested in the Heisenberg 
evolution.  Using the Dyson series for $\tilde U_t(\varepsilon)$, we now 
expand $\langle\psi_1|\tilde U_t(\varepsilon)^\dag X 
\tilde U_t(\varepsilon)\psi_2\rangle$ as
\begin{multline*}
	\sum_{n=0}^\infty\sum_{m=0}^\infty
 	(-i)^{n-m} 
	\int_{\Delta_{m}(t)} dt_{m}\cdots dt_{1}
	\int_{\Delta_{n}(t)} ds_{n}\cdots ds_{1} \\
	\times\langle\psi_{1}|
	\tilde{\Upsilon}_{t_{1}}(\varepsilon)\cdots 
	\tilde{\Upsilon}_{t_{m}}(\varepsilon) X
	\tilde{\Upsilon}_{s_{n}}(\varepsilon)\cdots 
	\tilde{\Upsilon}_{s_{1}}(\varepsilon)
	\psi_{2}\rangle
\end{multline*}
(for notational simplicity, we write from this point on
$\int_{\Delta_0(t)}\cdots=I$).  This equals
\begin{multline*}
	\sum_{n=0}^\infty\sum_{m=0}^\infty
 	(-i)^{n-m} 
	\int_{\Delta_{m}(t)} dt_{m}\cdots dt_{1}
	\int_{\Delta_{n}(t)} ds_{n}\cdots ds_{1}
	\sum_{\mu_m\nu_m}\cdots\sum_{\mu_1\nu_1}
	\sum_{\alpha_n\beta_n}\cdots\sum_{\alpha_1\beta_1}
	\\
	\times\langle\psi_{1}|
	\check{E}_{\mu_{1}\nu_{1}}(t_{1},\varepsilon) 
	\cdots \check{E}_{\mu_{m}\nu_{m}}(t_{m},\varepsilon)
	X\check{E}_{\alpha_{n}\beta_{n}}(s_{n},\varepsilon) 
	\cdots \check{E}_{\alpha_{1}\beta_{1}}(s_{1},\varepsilon)
	\psi_{2}\rangle \\
	\times
	\langle\Phi|
	[\tilde{a}_{t_{1}}(\varepsilon)^{\dag}]^{\mu_{1}}
	[\tilde{a}_{t_{1}}(\varepsilon)]^{\nu_{1}}
	\cdots
	[\tilde{a}_{t_{m}}(\varepsilon)^{\dag}]^{\mu_{m}}
	[\tilde{a}_{t_{m}}(\varepsilon)]^{\nu_{m}}
	\\ \times
	[\tilde{a}_{s_{n}}(\varepsilon)^{\dag}]^{\alpha_{n}}
	[\tilde{a}_{s_{n}}(\varepsilon)]^{\beta_{n}}
	\cdots
	[\tilde{a}_{s_{1}}(\varepsilon)^{\dag}]^{\alpha_{1}}
	[\tilde{a}_{s_{1}}(\varepsilon)]^{\beta_{1}}
	\,\Phi\rangle,
\end{multline*}
where we have applied lemma \ref{lem:commutethrough}.  As before, we can 
use Wick's lemma to evaluate the vacuum matrix element.  Drawing vertices 
on a line in the correct order, assigning incoming and outgoing lines 
according to $\alpha,\beta,\mu,\nu$, and connecting them up, allows us to 
represent the vacuum matrix element as a sum over the usual diagrams. For 
example, a possible diagram in this case might be:
\vskip.5cm
\begin{center}
\setlength{\unitlength}{.05cm}
\begin{picture}(210,20)
\put(20,10){\circle*{2}}
\put(30,10){\circle*{2}}
\put(40,10){\circle*{2}}
\put(50,10){\circle*{2}}
\put(60,10){\circle*{2}}
\put(70,10){\circle*{2}}
\put(80,10){\circle*{2}}
\put(90,10){\circle*{2}}
\put(100,10){\circle*{2}}
\put(30,10){\oval(20,20)[t]}
\put(40,10){\oval(20,20)[t]}
\put(60,10){\oval(20,20)[t]}
\put(90,10){\oval(20,20)[t]}
\put(18,1){$t_{1}$}
\put(28,1){$t_{2}$}
\put(38,1){$t_{3}$}
\put(48,1){$t_{4}$}
\put(58,1){$t_{5}$}
\put(68,1){$t_{6}$}
\put(78,1){$t_{7}$}
\put(88,1){$t_{8}$}
\put(98,1){$t_{9}$}
\put(10,10){\dashbox{0.5}(100,0){ }}
\put(120,10){\dashbox{0.5}(80,0){ }}
\put(130,10){\circle*{2}}
\put(140,10){\circle*{2}}
\put(150,10){\circle*{2}}
\put(160,10){\circle*{2}}
\put(170,10){\circle*{2}}
\put(180,10){\circle*{2}}
\put(190,10){\circle*{2}}
\put(128,1){$s_{7}$}
\put(138,1){$s_{6}$}
\put(148,1){$s_{5}$}
\put(158,1){$s_{4}$}
\put(168,1){$s_{3}$}
\put(178,1){$s_{2}$}
\put(188,1){$s_{1}$}
\put(120,10){\oval(40,20)[t]}
%\put(140,10){\oval(20,20)[t]}
\put(175,10){\oval(10,10)[t]}
\put(175,10){\oval(30,20)[t]}
\end{picture}
\end{center}
%\vskip.25cm
\noindent
Note that we do not need to worry about the time ordering (which is 
obviously not satisfied in this case), as the commutators between $\tilde 
a_s(\varepsilon)$ and $\tilde a_t(\varepsilon)^\dag$ are symmetric in 
$s,t$; hence only the order in which the $\tilde a$'s and $\tilde 
a^\dag$'s occur will matter, and we can expand in terms of pair 
contractions in the usual way.

The first question that needs to be resolved is whether we still have
uniform control on the convergence of the Dyson series.  This does turn 
out to be the case.  The argument used previously to obtain the required 
estimates can be generalized also to the Heisenberg evolution, though the 
details of the argument are somewhat more involved in this case.  We refer 
to \cite[section 9]{Gou05} for further details.

The next problem is to determine which diagrams survive in the 
$\varepsilon\to 0^+$ limit.  It is not difficult to see that diagrams with 
contractions between $s$-variables which are not time-consecutive or 
between $t$-variables which are not time-consecutive will vanish in the 
limit; this follows directly from the previous arguments.  Hence all 
surviving diagrams must have only time-consecutive contractions within the 
$s$- and $t$-blocks.  On the other hand, note that we are not integrating 
over the simplex $\Delta_{m+n}(t)$, but rather over the product of 
simplices $\Delta_m(t)\times\Delta_n(t)$.  Therefore contractions between 
$s$- and $t$-variables do not necessarily give vanishing contributions,
provided that the corresponding lines in the diagram do not cross---in the 
latter case the contraction would force $s_i=t_l$ and $s_j=t_k$ in the 
limit $\varepsilon\to 0^+$, whereas integration over $\Delta_m(t)\times
\Delta_n(t)$ requires $s_i<s_j$ and $t_k<t_l$.  For example,
\vskip.5cm
\begin{center}
\setlength{\unitlength}{.05cm}
\begin{picture}(210,20)
\put(20,10){\circle*{2}}
\put(30,10){\circle*{2}}
\put(40,10){\circle*{2}}
\put(50,10){\circle*{2}}
\put(60,10){\circle*{2}}
\put(70,10){\circle*{2}}
\put(80,10){\circle*{2}}
\put(90,10){\circle*{2}}
\put(100,10){\circle*{2}}
\put(25,10){\oval(10,10)[t]}
\put(35,10){\oval(10,10)[t]}
\put(95,10){\oval(10,10)[t]}
\put(18,1){$t_{1}$}
\put(28,1){$t_{2}$}
\put(38,1){$t_{3}$}
\put(48,1){$t_{4}$}
\put(58,1){$t_{5}$}
\put(68,1){$t_{6}$}
\put(78,1){$t_{7}$}
\put(88,1){$t_{8}$}
\put(98,1){$t_{9}$}
\put(10,10){\dashbox{0.5}(100,0){ }}
\put(120,10){\dashbox{0.5}(80,0){ }}
\put(130,10){\circle*{2}}
\put(140,10){\circle*{2}}
\put(150,10){\circle*{2}}
\put(160,10){\circle*{2}}
\put(170,10){\circle*{2}}
\put(180,10){\circle*{2}}
\put(190,10){\circle*{2}}
\put(128,1){$s_{7}$}
\put(138,1){$s_{6}$}
\put(148,1){$s_{5}$}
\put(158,1){$s_{4}$}
\put(168,1){$s_{3}$}
\put(178,1){$s_{2}$}
\put(188,1){$s_{1}$}
\put(125,10){\oval(50,20)[t]}
\put(155,10){\oval(10,10)[t]}
\put(165,10){\oval(10,10)[t]}
\put(175,10){\oval(10,10)[t]}
\put(135,10){\oval(10,10)[t]}
\put(100,10){\oval(60,15)[t]}
\end{picture}
\end{center}
%\vskip.25cm blibla
\noindent
must necessarily vanish, whereas the diagram
\vskip.5cm
\begin{center}
\setlength{\unitlength}{.05cm}
\begin{picture}(210,20)
\put(20,10){\circle*{2}}
\put(30,10){\circle*{2}}
\put(40,10){\circle*{2}}
\put(50,10){\circle*{2}}
\put(60,10){\circle*{2}}
\put(70,10){\circle*{2}}
\put(80,10){\circle*{2}}
\put(90,10){\circle*{2}}
\put(100,10){\circle*{2}}
\put(25,10){\oval(10,10)[t]}
\put(35,10){\oval(10,10)[t]}
\put(95,10){\oval(10,10)[t]}
\put(18,1){$t_{1}$}
\put(28,1){$t_{2}$}
\put(38,1){$t_{3}$}
\put(48,1){$t_{4}$}
\put(58,1){$t_{5}$}
\put(68,1){$t_{6}$}
\put(78,1){$t_{7}$}
\put(88,1){$t_{8}$}
\put(98,1){$t_{9}$}
\put(10,10){\dashbox{0.5}(100,0){ }}
\put(120,10){\dashbox{0.5}(80,0){ }}
\put(130,10){\circle*{2}}
\put(140,10){\circle*{2}}
\put(150,10){\circle*{2}}
\put(160,10){\circle*{2}}
\put(170,10){\circle*{2}}
\put(180,10){\circle*{2}}
\put(190,10){\circle*{2}}
\put(128,1){$s_{7}$}
\put(138,1){$s_{6}$}
\put(148,1){$s_{5}$}
\put(158,1){$s_{4}$}
\put(168,1){$s_{3}$}
\put(178,1){$s_{2}$}
\put(188,1){$s_{1}$}
\put(95,10){\oval(110,20)[t]}
\put(155,10){\oval(10,10)[t]}
\put(165,10){\oval(10,10)[t]}
\put(175,10){\oval(10,10)[t]}
\put(135,10){\oval(10,10)[t]}
\put(100,10){\oval(60,15)[t]}
\end{picture}
\end{center}
%\vskip.25cm
\noindent
could give a nonvanishing contribution to the Dyson expansion.  To
characterize such diagrams, we begin as before by specifying in increasing
time order the numbers $r_1,\ldots,r_p$ of vertices connected through
contractions within the $s$-block, and specifying the numbers
$l_1,\ldots,l_q$ of vertices connected through contractions within the
$t$-block, also in increasing time order.  For example, the nonvanishing
diagram above is described by the sequences $r=(1,4,2)$ and
$l=(3,1,1,1,1,2)$.  This specifies completely the (time-consecutive)
contractions within the $s$- and $t$-blocks.

It remains to specify the contractions between $s$- and $t$-variables.  
Note that we can only get additional contractions between the left
endpoint of a connected component in the $s$-block with the right endpoint
of a connected component in the $t$-block.  Let us write $\kappa_i=1$ if
the (left endpoint of the) $i$th connected component in the $s$-block is
contracted with a vertex in the $t$-block, and $\kappa_i=0$ otherwise;
similarly, we write $\lambda_i=1$ if the (right endpoint of the) $i$th
connected component in the $t$-block is contracted with a vertex in the
$s$-block, and $\lambda_i=0$ otherwise (note that necessarily 
$\sum\kappa=\sum\lambda$).  For example, the nonvanishing 
diagram above is described by $\kappa=(0,1,1)$ and $\lambda=(1,0,0,1,0,0)$.
Finally, we denote by $\kappa(i)$ the $i$th nonzero element of $\kappa$,
and similarly for $\lambda(i)$.  For example, in the nonvanishing 
diagram above, $\kappa(1)=2$, $\kappa(2)=3$, and $\lambda(1)=1$, 
$\lambda(2)=4$.  Once we have given $r$, $l$, $\kappa$ and $\lambda$ we 
have described uniquely one nonvanishing diagram, as the order in which 
the $s$--$t$ contractions are made is fixed by the requirement that the 
corresponding lines be noncrossing (we must connect the lines from the 
inside out, i.e.\ connected component $\kappa(i)$ is contracted with
connected component $\lambda(i)$).

With this somewhat tedious notation, we can write out the limiting Dyson 
series explicitly.  Applying Wick's lemma, retaining only the nonvanishing 
diagrams, and taking the limit as $\varepsilon\to 0^+$ gives
\begin{multline*}
	\sum_{n=0}^\infty\sum_{m=0}^\infty
	\sum_{\hat n,\hat m}
	\sum_{r_1,\ldots,r_{\hat n}}^{\sum r=n}
	\sum_{l_1,\ldots,l_{\hat m}}^{\sum l=m}
	\frac{(-i)^{n-m}}{2^{n-\hat n}2^{m-\hat m}}
	\sum_{\kappa,\lambda}^{\sum\kappa=\sum\lambda}
	\int_{\Delta_{\hat m}(t)} dt_{\hat m}\cdots dt_{1}
	\int_{\Delta_{\hat n}(t)} ds_{\hat n}\cdots ds_{1}
	\\
	\times\langle\psi_{1}|
	\check{E}_{0\lambda_{1}}^{(l_1)}(t_{1}) 
	\cdots \check{E}_{0\lambda_{\hat m}}^{(l_{\hat m})}(t_{\hat m})
	X\check{E}_{\kappa_{\hat n}0}^{(r_{\hat n})}(s_{\hat n}) 
	\cdots \check{E}_{\kappa_{1}0}^{(r_1)}(s_{1})
	\psi_{2}\rangle
	\,\prod_i\delta(s_{\kappa(i)}-t_{\lambda(i)}),
\end{multline*}
where we have written
$$
	\check{E}^{(r)}_{\alpha\beta}(t)=\left\{
	\begin{array}{ll}
		\check{E}_{\alpha\beta}(t) & \quad r=1, \\
		\check{E}_{\alpha 1}(t)(\check{E}_{11}(t))^{r-2}
		\check{E}_{1\beta}(t) & \quad r\ge 2.
	\end{array}
	\right.
$$
We could proceed at this point to resum the Dyson series as before, but 
instead it will be more convenient to work backwards from the desired 
result and show that we can recover the expression above.

Consider once more the Hudson-Parthasarathy equation
$$
	d\tilde U_t=\left\{
		L_{11}\,d\Lambda_t+
		L_{10}\,dA_t^\dag+
		L_{01}\,dA_t+
		L_{00}\,dt
	\right\}\tilde U_t.
$$
We are interested in the matrix element 
$$
	\langle\psi_1|\tilde U_t^\dag X\tilde U_t\,\psi_2\rangle=
	\langle\alpha_1|\alpha_2\rangle_{\rm osc}
	\langle v_1\otimes\Phi|
		e^{A(f_{1})}\,
		\tilde U_t^\dag X\tilde U_t\,
		e^{A(f_{2})^\dag}
	\,v_2\otimes\Phi\rangle.
$$
Using the It\^o rules, we can commute the field operators past the 
unitaries; then
$$
	\langle\psi_1|\tilde U_t^\dag X\tilde U_t\,\psi_2\rangle=
	\langle\alpha_1|\alpha_2\rangle_{\rm osc}
	\langle f_1|f_2\rangle_{\rm resv}
	\langle v_1\otimes\Phi|
		\check U_t^+ X\check U_t
	\,v_2\otimes\Phi\rangle,
$$
where we have written
\begin{eqnarray*}
	d\check U_t &=& \left\{
		\check L_{11}(t)\,d\Lambda_t+
		\check L_{10}(t)\,dA_t^\dag+
		\check L_{01}(t)\,dA_t+
		\check L_{00}(t)\,dt
	\right\}\check U_t,
\\
	d\check U_t^+ &=& \check U_t^+\left\{
		\check L_{11}^+(t)\,d\Lambda_t+
		\check L_{10}^+(t)\,dA_t^\dag+
		\check L_{01}^+(t)\,dA_t+
		\check L_{00}^+(t)\,dt
	\right\},
\end{eqnarray*}
and where the coefficients are given by
\begin{eqnarray*}
	\check L_{11}(t) &=& L_{11},\quad
	\check L_{10}(t)=L_{10}+L_{11}f_2(t),\quad
	\check L_{01}(t)=L_{01}+f_1(t)^*L_{11},
\\
	\check L_{11}^+(t) &=& L_{11}^\dag,\quad
	\check L_{10}^+(t)=L_{01}^\dag+L_{11}^\dag f_2(t),\quad
	\check L_{01}^+(t)=L_{10}^\dag+f_1(t)^*L_{11}^\dag,
\end{eqnarray*}	
and
$$
	\check L_{00}(t)=\sum_{\alpha\beta}[f_1(t)^*]^\alpha 
		L_{\alpha\beta}[f_2(t)]^\beta,\quad
	\check L_{00}^+(t)=\sum_{\alpha\beta}[f_1(t)^*]^\alpha 
		L_{\beta\alpha}^\dag[f_2(t)]^\beta.
$$
But by explicit summation one may verify that
$$
	\check L_{\alpha\beta}(t)=
	\sum_{r\ge 1}\frac{\check E_{\alpha\beta}^{(r)}(t)}{i^r2^{r-1}},
	\qquad
	\check L_{\alpha\beta}^+(t)=
	\sum_{r\ge 1}\frac{\check E_{\alpha\beta}^{(r)}(t)}{(-i)^r2^{r-1}}.
$$
Using Picard iteration to develop $\check U_t$ and $\check U_t^+$ into 
their chaotic expansions, substituting the above expressions for $\check 
L,\check L^+$ and rearranging the summations somewhat, we arrive at the 
following Dyson expansion for  $\langle v_1\otimes\Phi|\check U_t^+ 
X\check U_t\,v_2\otimes\Phi\rangle$:
\begin{multline*}
	\sum_{n=0}^\infty\sum_{m=0}^\infty
	\sum_{\hat n,\hat m}
	\sum_{r_1,\ldots,r_{\hat n}}^{\sum r=n}
	\sum_{l_1,\ldots,l_{\hat m}}^{\sum l=m}
	\frac{(-i)^{n-m}}{2^{n-\hat n}2^{m-\hat m}}
	\sum_{\alpha_{\hat n}\beta_{\hat n}}\cdots
	\sum_{\alpha_1\beta_1}
	\sum_{\mu_{\hat m}\nu_{\hat m}}\cdots
	\sum_{\mu_1\nu_1} \\
	\times\langle v_1\otimes\Phi|
	\int_{\Delta_{\hat m}(t)}
		\check{E}^{(l_1)}_{\mu_1\nu_1}(t_1)\cdots
		\check{E}^{(l_{\hat m})}_{\mu_{\hat m}\nu_{\hat m}}(t_{\hat m})
		\,
		d\Lambda^{\mu_{\hat m}\nu_{\hat m}}_{t_{\hat m}}\cdots
		d\Lambda^{\mu_1\nu_1}_{t_1}
	\\
	\times X \times
	\int_{\Delta_{\hat n}(t)}
		\check{E}^{(r_{\hat n})}_{\alpha_{\hat n}\beta_{\hat n}}(s_{\hat n})
		\cdots
		\check{E}^{(r_1)}_{\alpha_1\beta_1}(s_1)
		\,
		d\Lambda^{\alpha_{\hat n}\beta_{\hat n}}_{s_{\hat n}}\cdots
		d\Lambda^{\alpha_1\beta_1}_{s_1}
	\,v_2\otimes\Phi\rangle.
\end{multline*}
Using the quantum It\^o rules and by induction on the iterated integrals, 
it is not difficult to establish that the vacuum matrix element in this 
expression vanishes if any of the $\mu_i$ or $\beta_i$ are nonzero, or if 
the number of nonzero $\nu$'s and $\alpha$'s do not coincide.  Hence we 
find, relabeling the variables suggestively,
\begin{multline*}
	\sum_{n=0}^\infty\sum_{m=0}^\infty
	\sum_{\hat n,\hat m}
	\sum_{r_1,\ldots,r_{\hat n}}^{\sum r=n}
	\sum_{l_1,\ldots,l_{\hat m}}^{\sum l=m}
	\frac{(-i)^{n-m}}{2^{n-\hat n}2^{m-\hat m}}
	\sum_{\kappa,\lambda}^{\sum\kappa=\sum\lambda}
	\\
	\times\langle v_1\otimes\Phi|
	\int_{\Delta_{\hat m}(t)}
		\check{E}^{(l_1)}_{0\lambda_1}(t_1)\cdots
		\check{E}^{(l_{\hat m})}_{0\lambda_{\hat m}}(t_{\hat m})
		\,
		d\Lambda^{0\lambda_{\hat m}}_{t_{\hat m}}\cdots
		d\Lambda^{0\lambda_1}_{t_1}
	\\
	\times X \times
	\int_{\Delta_{\hat n}(t)}
		\check{E}^{(r_{\hat n})}_{\kappa_{\hat n}0}(s_{\hat n})
		\cdots
		\check{E}^{(r_1)}_{\kappa_1 0}(s_1)
		\,
		d\Lambda^{\kappa_{\hat n}0}_{s_{\hat n}}\cdots
		d\Lambda^{\kappa_1 0}_{s_1}
	\,v_2\otimes\Phi\rangle.
\end{multline*}
But now we can easily reduce to the previous form of the Dyson expansion, 
taking into account the identity (which follows directly from the
quantum It\^o rules)
\begin{multline*}
	\langle v\otimes\Phi|
	\int_0^tF_\tau\,dA_\tau\times
	\int_0^sG_\sigma\,dA_\sigma^\dag\,
	w\otimes\Phi\rangle \\ =
	\int_0^t d\tau\int_0^s d\sigma\,
		\langle v\otimes\Phi|
		F_\tau G_\sigma\,w\otimes\Phi\rangle\,
		\delta(\tau-\sigma).
\end{multline*}
The proof of Theorem 2 is complete.

\subsection{Proof of Theorem \ref{thm:mainfield}}

The hard work has already been done in the proof of Theorem 
\ref{thm:mainatom}; all we have to do to prove Theorem \ref{thm:mainfield} 
is an appropriate shift of the coefficients.  We briefly provide the 
details.  Consider first the expansion for $\langle\psi_1|U_t(\varepsilon)^\dag
W(g_{t]})U_t(\varepsilon)\psi_2\rangle$,
\begin{multline*}
	\sum_{n=0}^\infty\sum_{m=0}^\infty
 	(-i)^{n-m} 
	\int_{\Delta_{m}(t)} dt_{m}\cdots dt_{1}
	\int_{\Delta_{n}(t)} ds_{n}\cdots ds_{1}
	\times\langle v_{1}\otimes\Phi|
	\check{\Upsilon}_{t_{1}}(\varepsilon)\cdots 
	\check{\Upsilon}_{t_{m}}(\varepsilon) \\ \times
		W(g_{t]}-2g_{t],\varepsilon}^+)
		B(\sqrt{\tfrac{4\varepsilon}{\gamma}}\,g_{t]}^+(0,\varepsilon))
	\check{\Upsilon}_{s_{n}}(\varepsilon)\cdots 
	\check{\Upsilon}_{s_{1}}(\varepsilon)\,
	v_2\otimes\Phi\rangle,
\end{multline*}
where we have dropped the prefactor $\langle\alpha_1|\alpha_2\rangle_{\rm osc}
\,\langle f_1|f_2\rangle_{\rm resv}$ and the constant factor that is 
obtained from commuting $e^{A(f_{1})}$, etc., past the Weyl operators.
Splitting up the Weyl operators as explained in section \ref{sec:output} 
and commuting them through the Hamiltonians $\check{\Upsilon}$ as in lemma 
\ref{lem:commutethrough} gives
\begin{multline*}
	\sum_{n=0}^\infty\sum_{m=0}^\infty
 	(-i)^{n-m} 
	\int_{\Delta_{m}(t)} dt_{m}\cdots dt_{1}
	\int_{\Delta_{n}(t)} ds_{n}\cdots ds_{1} \\
	\times\langle v_{1}\otimes\Phi|
	\check{\Upsilon}^\wedge_{t_{1}}(\varepsilon)\cdots 
	\check{\Upsilon}^\wedge_{t_{m}}(\varepsilon)
	\check{\Upsilon}^\vee_{s_{n}}(\varepsilon)\cdots 
	\check{\Upsilon}^\vee_{s_{1}}(\varepsilon)
	\,v_{2}\otimes\Phi\rangle,
\end{multline*}
where we have dropped the constant factor that is obtained when we split 
the Weyl operators.  Here $\check{\Upsilon}^\wedge_{s}$ is obtained from 
$\check{\Upsilon}_s$ by transforming
\begin{eqnarray*}
&&	\check{E}_{10}(s,\varepsilon)\mapsto
	\check{E}_{10}(s,\varepsilon)+
	\left[\{g_{t]}-2g_{t],\varepsilon}^+\}^-(s,\varepsilon)
	-\frac{4\varepsilon}{\gamma}\,g_{t]}^+(0,\varepsilon)\,G_\varepsilon(s)
	\right]\check{E}_{11}(s,\varepsilon)
	,\\
&&	\check{E}_{00}(s,\varepsilon)\mapsto
	\check{E}_{00}(s,\varepsilon)+
	\left[\{g_{t]}-2g_{t],\varepsilon}^+\}^-(s,\varepsilon)
	-\frac{4\varepsilon}{\gamma}\,g_{t]}^+(0,\varepsilon)\,G_\varepsilon(s)
	\right]\check{E}_{01}(s,\varepsilon)
	,
\end{eqnarray*}
and $\check{\Upsilon}^\vee_{s}$ is obtained from $\check{\Upsilon}_s$ by 
transforming
\begin{eqnarray*}
&&	\check{E}_{01}(s,\varepsilon)\mapsto
	\check{E}_{01}(s,\varepsilon)-
	\left[\{g_{t]}-2g_{t],\varepsilon}^+\}^{-}(s,\varepsilon)^*
	-\frac{4\varepsilon}{\gamma}\,g_{t]}^+(0,\varepsilon)^*
	\,G_\varepsilon(s)
	\right]\check{E}_{11}(s,\varepsilon)
	,\\
&&	\check{E}_{00}(s,\varepsilon)\mapsto
	\check{E}_{00}(s,\varepsilon)-
	\left[\{g_{t]}-2g_{t],\varepsilon}^+\}^{-}(s,\varepsilon)^*
	-\frac{4\varepsilon}{\gamma}\,g_{t]}^+(0,\varepsilon)^*
	\,G_\varepsilon(s)
	\right]\check{E}_{10}(s,\varepsilon).
\end{eqnarray*}
We can now proceed exactly as in the proof of Theorem \ref{thm:mainatom} 
to establish that in the limit $\varepsilon\to 0^+$, this expansion 
reduces to
\begin{multline*}
	\sum_{n=0}^\infty\sum_{m=0}^\infty
	\sum_{\hat n,\hat m}
	\sum_{r_1,\ldots,r_{\hat n}}^{\sum r=n}
	\sum_{l_1,\ldots,l_{\hat m}}^{\sum l=m}
	\frac{(-i)^{n-m}}{2^{n-\hat n}2^{m-\hat m}}
	\sum_{\kappa,\lambda}^{\sum\kappa=\sum\lambda}
	\int_{\Delta_{\hat m}(t)} dt_{\hat m}\cdots dt_{1}
	\int_{\Delta_{\hat n}(t)} ds_{\hat n}\cdots ds_{1}
	\\
	\times\langle v_{1}|
	\check{E}_{0\lambda_{1}}^{(l_1)\wedge}(t_{1}) 
	\cdots \check{E}_{0\lambda_{\hat m}}^{(l_{\hat m})\wedge}(t_{\hat m})
	\check{E}_{\kappa_{\hat n}0}^{(r_{\hat n})\vee}(s_{\hat n}) 
	\cdots \check{E}_{\kappa_{1}0}^{(r_1)\vee}(s_{1})\,
	v_{2}\rangle
	\,\prod_i\delta(s_{\kappa(i)}-t_{\lambda(i)}),
\end{multline*}
where $\check{E}_{0\lambda}^{(l)\wedge}(s)$ is obtained through the 
replacements
$$
	\check{E}_{10}(s)\mapsto
	\check{E}_{10}(s)-g(s)\check{E}_{11}(s),\qquad
	\check{E}_{00}(s)\mapsto
	\check{E}_{00}(s)-g(s)\check{E}_{01}(s),
$$
and $\check{E}_{\kappa 0}^{(r)\vee}(s)$ is obtained through the 
replacements
$$
	\check{E}_{01}(s)\mapsto
	\check{E}_{01}(s)+g(s)^*\check{E}_{11}(s),\qquad
	\check{E}_{00}(s)\mapsto
	\check{E}_{00}(s)+g(s)^*\check{E}_{10}(s).
$$
Starting from the opposite direction, it is not difficult to establish 
that the expected result of Theorem \ref{thm:mainfield},
$\langle\psi_1|\tilde U_t^\dag W(-g_{t]})\tilde U_t\psi_2\rangle$, can be 
written (modulo prefactor) as $\langle v_1\otimes\Phi|\check{U}_t^\wedge
\check{U}_t^\vee\,v_2\otimes\Phi\rangle$, where $\check{U}_t^\wedge$ is 
obtained from $\check{U}_t^+$ by the replacements
$$
	\check L_{10}^+(s)\mapsto \check L_{10}^+(s)-g(s)
		\check L_{11}^+(s),\qquad
	\check L_{00}^+(s)\mapsto \check L_{00}^+(s)-g(s)
		\check L_{01}^+(s),
$$
and $\check{U}_t^\vee$ is obtained from $\check{U}_t$ by the replacements
$$
	\check L_{01}(s)\mapsto \check L_{01}(s)+g(s)^*\check L_{11}(s),\qquad
	\check L_{00}(s)\mapsto \check L_{00}(s)+g(s)^*\check L_{10}(s).
$$
It is important to note that the constant factor which we have dropped 
here is precisely the limit as $\varepsilon\to 0^+$ of the constant factor 
that was dropped previously; hence it suffices to show that the two 
expansions above coincide.  However, this is immediate from our previous 
results, and the theorem is proved.

\begin{acknowledgements}
The authors thank Luc Bouten and Hideo Mabuchi for insightful discussions.  
R.v.H. thanks Andrew Doherty and Howard Wiseman for their comments on the 
relation of this paper to previous work.
\end{acknowledgements}

\end{document}